\documentclass[lettersize,journal]{IEEEtran}

\usepackage{ifpdf}
\usepackage{cite}
\ifCLASSINFOpdf
\usepackage[pdftex]{graphicx}
\usepackage{amsmath}

\DeclareMathOperator*{\argmin}{arg\,min}
\usepackage{amsfonts} 
\usepackage{multirow}
\usepackage{algorithmic}
\usepackage{array}
\usepackage{caption}
\usepackage{subcaption}
\usepackage{cleveref}
\usepackage{mhchem}
\expandafter\def\expandafter\quote\expandafter{\quote\small}
\usepackage{float}

\usepackage{fixltx2e}
\usepackage{stfloats}
\usepackage{url}

\begin{document}

\title{A Multiobjective Reinforcement Learning Framework for Microgrid Energy Management}

\author{~M. Vivienne Liu,
        ~Patrick M. Reed,
        ~David Gold,
        ~Garret Quist,
        and~C. Lindsay Anderson,~\IEEEmembership{Senior Member,~IEEE}

\thanks{M. V. Liu is with the Field of Systems Engineering, Cornell University, Ithaca, NY, 14853, USA (e-mail: ml2589@cornell.edu)}
\thanks{P. M. Reed and D. Gold are with Civil and Environmental Engineering, Cornell University, Ithaca, NY, 14853, USA (e-mail: patrick.reed@cornell.edu; dfg42@cornell.edu).}
\thanks{G. Quist is with the Facilities and Campus Service, Cornell University, Ithaca, NY, 14853, USA (e-mail: gq29@cornell.edu).}
\thanks{C. L. Anderson is with the Field of Systems Engineering and Cornell Energy Systems Institute, Cornell University, Ithaca, NY, 14853, USA (e-mail: cla28@cornell.edu)}
}



\maketitle
\begin{abstract}
The emergence of microgrids (MGs) has provided a promising solution for decarbonizing and decentralizing the power grid, mitigating the challenges posed by climate change. However, MG operations often involve considering multiple objectives that represent the interests of different stakeholders, leading to potentially complex conflicts. To tackle this issue, we propose a novel multi-objective reinforcement learning framework that explores the high-dimensional objective space and uncovers the tradeoffs between conflicting objectives. This framework leverages exogenous information and capitalizes on the data-driven nature of reinforcement learning, enabling the training of a parametric policy without the need for long-term forecasts or knowledge of the underlying uncertainty distribution. The trained policies exhibit diverse, adaptive, and coordinative behaviors with the added benefit of providing interpretable insights on the dynamics of their information use. We employ this framework on the Cornell University MG (CU-MG), which is a combined heat and power MG, to evaluate its effectiveness. The results demonstrate performance improvements in all objectives considered compared to the status quo operations and offer more flexibility in navigating complex operational tradeoffs.
\end{abstract}


\section{Introduction} \label{sec: introduction}
As the urgency to reduce greenhouse gas emissions grows, Microgrids(MGs) are emerging as an effective and efficient solution for integrating distributed renewable resources~\cite{lasseter2002microgrids}. Compared to centralized power grids, MGs leverage localized generation, minimizing transmission losses and eliminating the need for significant infrastructure adjustments. Another noteworthy advantage of MGs is their ability to enhance grid resilience by operating in ``islanded mode," enabling self-sustainability during outages~\cite{majzoobi2016application}. Furthermore, MGs play a crucial role in electrifying rural areas, leading to a substantial surge in their numbers~\cite{zhou2015overview}.

The energy management of a MG, involving distributed generation units (both conventional and renewable) and energy storage units, requires making sequential decisions in the face of uncertainties introduced by renewable energy sources. The optimal dispatch for MG energy management under these uncertainties is often addressed by representing the control problem as a Markov Decision Process (MDP)~\cite{stvrelec2012modeling,wu2013coordinated}. The MDP represents a sequential decision process in which an agent interacts with an environment, making decisions at each time step to maximize its long-term expected rewards~\cite{puterman1990markov}. Traditionally, the MDP control formulations have been tackled using the Dynamic Programming (DP) methods, which are often burdened by the ``curse of dimensionality," limiting the size of the state and action spaces that can be addressed~\cite{zeng2018dynamic}. Moreover, traditional MDP formulations require knowledge of the ``model", which refers to the transition probability and reward function for given state-action pairs for the system of consideration. Unfortunately, in energy systems these transition probabilities are often not known due to the complexity and non-linearity of the systems, prompting the need for a ``model-free" approach.

To overcome these issues, Reinforcement Learning (RL) has gained popularity in power system operation and control due to its model-free nature~\cite{zhang2019deep}. RL, also known as model-free approximate dynamic programming (ADP), has been combined with artificial neural networks (ANNs) for function approximations, leading to the development of Deep RL (DRL) methods~\cite{mousavi2018deep}. Numerous advancements have been made in employing (D)RL algorithms to enhance demand-side energy management~\cite{zeng2018dynamic,zhang2019learning,liu2018distributed}. For a comprehensive overview of RL in the context of power system operations, interested readers can refer to~\cite{chen2022reinforcement}. A common practice of these studies is to take advantage of the data-driven property of RL and use historical data to train the agent by simulating its interaction with the environment without oversimplifying the state and action spaces. However, prior studies have primarily focused on a single objective, typically maximizing profit (minimizing cost), due to the inherent nature of the underlying Bellman equations formulation. In fact, methods focused on multiple objectives are exceedingly rare in the RL and ADP bodies of literature~\cite{powell2019unified,bertsekas2019reinforcement}. As the share of renewable energy increases in MGs, other objectives, such as environmental impacts, operation reliability, and effective storage operation, are receiving more attention~\cite{khan2016compendium}. A sole focus on a single objective could lead to control solutions in the extreme corners of the broader space of relevant performance objectives and fails to properly represent the interests of stakeholders, especially when conflicts exist between their objective functions~\cite{woodruff2013many,gupta2020exploring}.

In addressing multiple objectives in MG optimization, a standard approach is to employ a weighted sum method to convert the original multiple objective formulations into a single objective representation that tacitly infers that the specified weights capture all stakeholders' preferences. The weighted single objective is the dominant approach for handling multiple objectives in  DP formulations~\cite{castelletti2013multiobjective}. If a user is interested in an explicit representation of the full suite of objective tradeoffs, solving repeated weighted DP instances can become computationally prohibitive. Identifying all of the Pareto optimal solutions\footnote{Pareto-optimal solution refers to a solution where no objective can be improved without sacrificing another objective.} that compose control tradeoffs becomes increasingly challenging as the number of objectives grows, leading to a factorial growth in the computation cost~\cite{giuliani2021state}. In~\cite{yang2021multi,pourghasem2019stochastic,saberi2019optimal,conti2012optimal}, the weighted sum approach is combined with fuzzy techniques to limit the number of combinations of weights, reducing the complexity of the computation. However, these methods struggle to effectively explore the tradeoffs between the objectives. The predefined preferences for objectives can potentially overlook superior solutions, particularly in scenarios where system dynamics are non-convex and non-linear~\cite{salehi2022comprehensive}. This limitation hampers the ability to find optimal solutions that strike the best balance among conflicting objectives.

Meta-heuristic methods such as the Non-dominated Sorting Genetic Algorithm (NSGA-II)~\cite{deb2002fast} and Multiple Objective Particle Swarm Optimization (MOPSO)~\cite{coello2002mopso} have been employed in optimizing MG operation to allow for an explicit search of the Pareto-optimal solutions~\cite{teo2020optimization,elgammal2018energy}. These multi-objective optimization algorithms use population sorting techniques to guide search and maintain a set of non-dominated solutions, resulting in a diverse set of Pareto-optimal solutions for decision-makers without specifying predefined preferences for objectives. In addition, as simulation-based optimization, the meta-heuristic methods can handle more complex system formulations that are non-linear and non-convex, making them a good candidate class of solution tools for the demand-side energy management problem~\cite{WEITZEL2018582}.

In this paper, we present a novel framework that combines the strengths of multi-objective optimization and RL to tackle the energy management problem of MGs. We use the Borg Multi-Objective Evolutionary Algorithm (MOEA)~\cite{hadka2013borg}, which has been proven to meet or exceed the performance of other MOEAs in complex system planning and operation applications~\cite{reed2013evolutionary,gupta2020can,salazar2016diagnostic}, explicitly exploring their tradeoffs in the higher-dimensional objective spaces. Our framework proposes a model-free policy approximation approach, enabling the agent to interact with the unknown environment in continuous state and action spaces while managing the computation complexity of the stochastic MG control problem of focus. The proposed Multi-Objective Reinforcement Learning (MORL) framework is applied to the challenging abstraction of the Combined Heat and Power (CHP) CU-MG, demonstrating its efficacy in a real-world application. Using historical weather, demand, and utility price data, the trained agent (policy) makes collaborative decisions for multiple energy sources under stochastic state conditions. This model-free policy approximation approach not only reduces computational complexity but also provides interpretability in the trained parametric policy, allowing us to understand how the agent utilizes exogenous information to shape its dynamic and adaptive control actions. The primary contributions of this study are as follows:

\begin{itemize}
    \item a novel framework that combines the power of multi-objective optimization and RL to support MG energy management under uncertainty, 
    \item a data-driven model-free policy approximation that reduces computation complexity while providing interpretability to the policy trained under multiple conflicting objectives, and 
    \item demonstrated performance on an existing CHP MG to highlight the tradeoffs between multiple objectives and the importance of exogenous information in improving stochastic decisions relative to current operations. 
    \end{itemize}

The rest of this paper is organized as follows. Section~\ref{sec: MGmodel} introduces the real-world CU-MG and the system modeling formulations. Section~\ref{sec: method} presents the proposed MORL framework and formulates the multi-objective energy management problem under it. Section~\ref{sec: numericresults} demonstrates the numerical results for the test case. Section~\ref{sec: conclusion} concludes this paper.  

\section{Multi-objective Energy Management Formulation on CU-MG} \label{sec: MGmodel}

We use the CU-MG as a practical example to illustrate the effectiveness of the proposed MORL framework. With Cornell University's ambitious goal of achieving carbon neutrality by 2035~\cite{beckers2015hybrid}, the campus utility faces the challenge of balancing emission reduction with cost maintenance. Additionally, as a combined heat and power (CHP) MG, it is crucial to make coordinated decisions regarding power and steam generation to ensure both electricity and heating reliability. Consequently, the  CU-MG serves as an ideal test case to evaluate the efficacy of our framework.  In this section, we begin by introducing and modeling the key components of the MG. Subsequently, we formulate the objectives and constraints of the energy management problem.

\subsection{Introduction of the Microgrid} \label{sec: Microgridmodel}

\begin{figure}[h]
  \begin{center}
  \includegraphics[width=0.90\columnwidth]{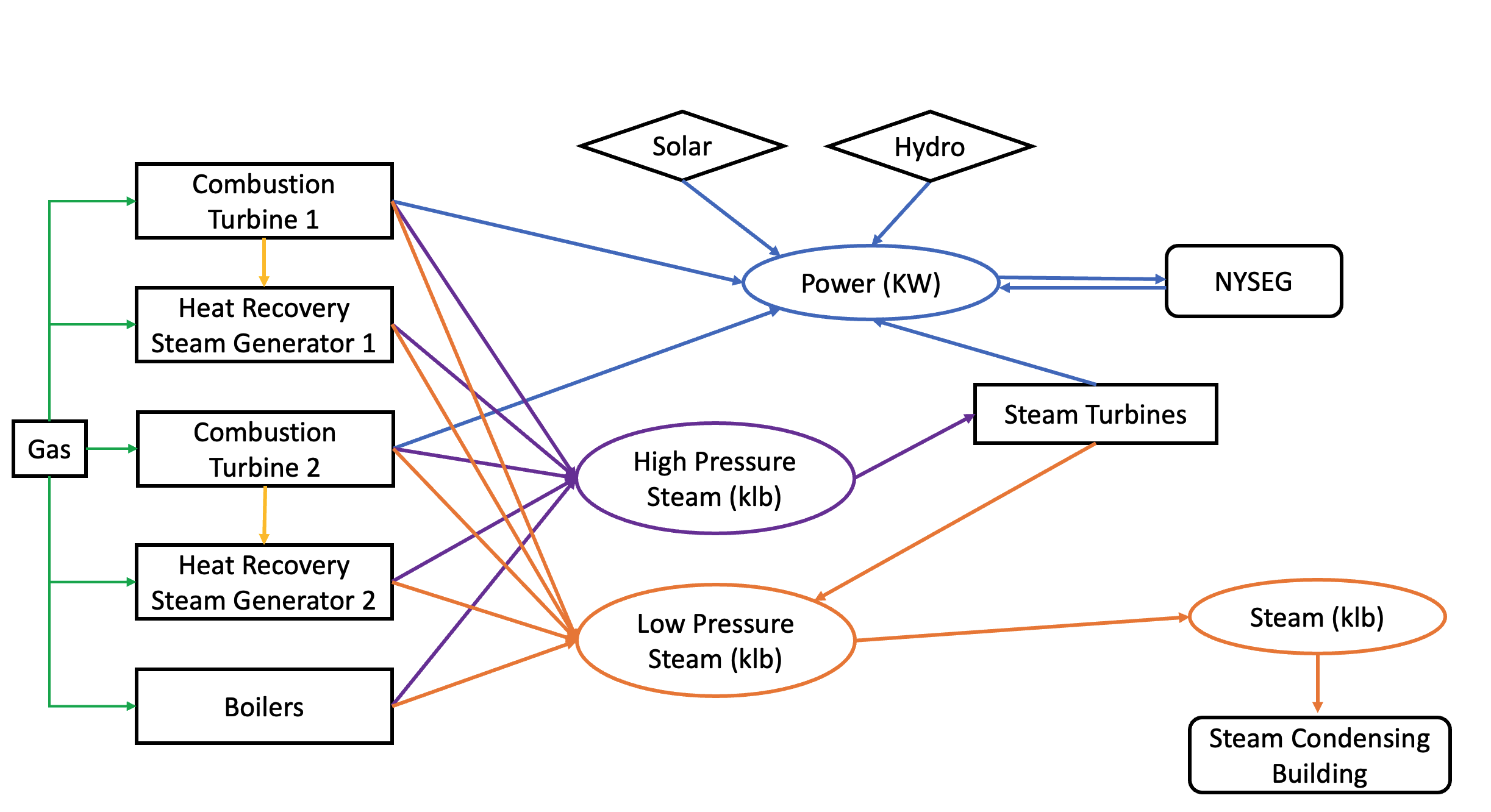}
  \caption{Overview of the CU-MG. }\label{fig: CornellMG}
    \vspace{-20pt}
  \end{center}
\end{figure}

The MG's combined heat and power (CHP) generators were commissioned in 2010, providing a capacity of 32 MW. The primary objectives behind their installation were to improve overall energy efficiency and decrease the campus carbon footprint. Each CHP unit consists of a Combustion Turbine (CT) and a Heat Recovery Steam Generator (HRSG), resulting in an impressive overall efficiency of approximately 80\% under optimal operating conditions. In 2012, the MG's energy input totaled 2,405,000 MMBtu, while the energy output amounted to 1,881,000 MMBtu. Notably, 62\% of this output was in the form of steam, while the remaining 38\% was in the form of electricity. The overview of the MG is summarized in Fig.~\ref{fig: CornellMG}.

The electric power supply in the MG, as depicted in Fig.~\ref{fig: CornellMG}, primarily relies on Combustion Turbines (CTs) fueled by natural gas. The Steam Turbine (ST) takes in high-pressure (400 psig) and high-temperature (600 F) steam to generate additional electricity, effectively harnessing "cogenerated" energy. Furthermore, the campus incorporates renewable energy sources, including approximately 0.2 MW of solar power and 1.5 MW of hydropower. The MG is also interconnected with the NYSEG utility grid at the Maple Ave substation through the Point of Common Coupling (PCC), allowing energy exchange between the MG and the utility grid.

Regarding the steam supply, the CTs are coupled with two Heat Recovery Steam Generators (HRSGs), which efficiently utilize the heat byproduct. These HRSGs enable the recovery of approximately 55 klb of steam without requiring additional natural gas input. Additionally, a set of boilers is responsible for generating both high-pressure (400 psig) and low-pressure (50 psig) steams for heating purposes across the campus. The high-pressure and high-temperature steam pass through the steam turbine generator to generate extra power before being converted to low-pressure steam for campus heating. The surplus steam is directed to the steam condenser building.

The components of the MG can be modeled as follows:
\begin{enumerate}
    \item \textbf{Combined Heat and Power Generator} 
    The CTs and HRSGs can be modeled as a CHP generator. The fuel cost of the CHP is then modeled as suggested by~\cite{aluisio2017optimization}:
    \begin{gather}
        C(p_{i,t,s}^c) = \dfrac{p_{i,t,s}^c}{H_i^c \cdot\eta(p_{i,t,s}^c)}\varphi_{t} \label{eq: chpeleccost}
    \end{gather}
    where $p_{i,t,s}^c$ is the electric power output for the ith CHP at hour $t$ in scenario $s$, $H_i^c$ is the heating value of the fuel, $\varphi_{t}$ is the price for natural gas and $\eta(p_{i,t,s}^c)$ is the electrical efficiency at output level $p_{i,t,s}^c$, which can be fitted as a function as follows given historical data:
    \begin{gather}
    \eta(p_{i,t,s}^c) = a_i^c + b_i^c \cdot (\dfrac{p_{i,t,s}^c}{\overline{P_i^c}}) + c_i^c \cdot (\dfrac{p_{i,t,s}^c}{\overline{P_i^c}})^2 \label{eq: chpeff}
    \end{gather}
    where $a_i^c$, $b_i^c$ and $c_i^c$ are coefficients fitted by historical data, $\overline{P_i^c}$ is the maximum capacity of the ith CHP. For details of the (fitted) parameters of the CHPs and other components in the MG, please refer to S.1 in Supplementary Information (SI). \\
    In addition, the output $p_{i,t,s}^c$ should satisfy the ramping constraint and upper/lower limits:
    \begin{gather}
    \underline{R_i^c} \leq p_{i,t,s}^c - p_{i,t-1,s}^c \leq \overline{R_i^c} \label{eq: chpramp}\\
        \underline{P_i^c} \leq p_{i,t,s}^c \leq \overline{P_i^c} \label{eq: chplim}
    \end{gather}
    The fuel cost of the steam output of the CHP depends on the volume of steam  generated. If the volume is less than approximately 55lb, no extra fuel is needed. Otherwise, the fuel needed can be fitted as a linear function of the steam output. The fuel cost can be modeled as:
    \begin{gather}
        C(q_{i,t,s}^c) =  \max(a_i^q \cdot q_{i,t,s}^c - b_i^q,0)\varphi_{t} \label{eq: chpsteamcost}
    \end{gather}
    where $a_i^q$ and $b_i^q$ are coefficients fitted by historical data. $q_{i,t,s}^c$ is the steam output of the ith CHP and it is bounded by the upper and lower bound $\overline{Q^c}$ and $\underline{Q_i^c}$:
    \begin{gather}
        \underline{Q_i^c} \leq q_{i,t,s}^c \leq \overline{Q_i^c} \label{eq: chpstlim}
    \end{gather}
    \item \textbf{Boilers}
    The fuel cost of the boiler can be formulated as
    \begin{gather}
        C(q_{t,s}^b) = (a^b + b^b \cdot (q_{t,s}^b) + c^b (q_{t,s}^b))^2)\varphi_{t} \label{eq: boilercost}
    \end{gather}
    where $a^c$, $b^c$ and $c^c$ are coefficients fitted by historical data. $q_{t,s}^b$ is the steam output of the boiler, which is constrained by its upper and lower bound $\overline{Q^b}$ and $ \underline{Q^b}$:
    \begin{gather}
        \underline{Q^b} \leq q_{t,s}^b \leq \overline{Q^b} \label{eq: boilerlim}
    \end{gather}
    \item \textbf{Steam Turbines}
    The ST generates additional electric power using the recovered steam and the relationship between the steam and output can be model as a piece-wise linear function as follows:
    \begin{gather}
        p_{t,s}^s =  \left\{ \begin{array}{rcl}
a_1^s \cdot (\sum_i q_{i,t,s}^c + q_{t,s}^b) + b_1^s & \mbox{if} \\ \sum_i q_{i,t,s}^c + q_{t,s}^b >c^s \\ 
a_2^s \cdot (\sum_i q_{i,t,s}^c + q_{t,s}^b) + b_2^s & \mbox{otherwise} &
\end{array}\right. \label{eq: stgen}
    \end{gather}
    where $a_1^s$, $b_1^s$, $a_2^s$, $b_2^s$ and $c^s$ are coefficients fitted by historical data. $p_{t,s}^s$ is the electricity output of the ST.
    \item \textbf{Power Exchange at PCC}
    The CHP MG can buy or sell power through the PCC and the cost/revenue can be modeled as:
    \begin{gather}
        C(p_{t,s}^e) = \pi_{t,s} \cdot p_{t,s}^e \label{eq: pcc}
    \end{gather}
    where $\pi_{t,s}$ is the real-time location marginal price of the Central load zone in New York Independent System Operator (NYISO) and $p_{t,s}^e$ is the energy exchange for hour $t$ in scenario $s$. $p_{t,s}^e>0$ indicates that buying energy from the utility and $p_{t,s}^e<0$ means selling energy to the utility. Currently, there is no limit on the energy exchange.
    \item \textbf{Renewable Inputs}
    The renewable energy output of hydro and solar are modeled as unobservant states of the system. It is assumed that the agent can only observe the realizations of these uncertain signals after the action is made. The available hydro and solar energy for hour $t$ under scenario $s$ are denoted by $y_{t,s}^h$ and $y_{t,s}^s$, respectively.
\end{enumerate}

\subsection{Formulation the energy management problem} \label{sec: opffomulation}
In this section, the MG energy management problem is presented as a multi-objective optimization problem.

\subsubsection{Objective Functions}
Three objective functions are selected to optimize the framework: the expected cost, expected emission, and expected heat waste. While selection of the first two objectives is straightforward, we note that the last objective is chosen to improve overall efficiency. This is particularly relevant in summer when the requirement for steam is significantly lower than in winter. Note that when the steam byproduct cannot be used, the overall efficiency of the CHPs decreases. These objective functions can be modeled as follows.

\textbf{Expected Cost}: The expected cost is made up of the fuel cost of local generation and the cost/revenue for energy exchange with the utility grid. The expected cost over all the scenarios $S$ can be formulated as:
    \begin{equation}
    \begin{aligned}
        O_1 = \dfrac{1}{|S|} \sum_{s = 1}^S &\sum_{t=1}^T \Biggl(\sum_{i=1}^{nc} \left(C(p_{i,t,s}^c) +C(q_{i,t,s}^c)\right)\\
        &+C(q_{t,s}^b)+C(p_{t,s}^e)\Biggr) \label{eq: costobj}
    \end{aligned}
    \end{equation}
    
\textbf{Expected Emission}: The expected emission consists of those from the local MG and from electricity provided by NYSEG. The emission of the local grid is based on the equivalent emission factor, which can be written as:
    \begin{gather}
        EF_{s,t}^{eq} = \dfrac{g_{t,s} \cdot \epsilon}{L_{t,s}+HL_{t,s}}
    \end{gather}
where $\epsilon$ is the emission rate of natural gas, $L_{t,s}$ and $HL_{t,s}$ are the electric and heat load of hour $t$ for scenario $s$, respectively. $g_{t,s}$ is the total natural gas consumed by the system, which can be calculated by:
    \begin{equation}
        \begin{aligned}
        g_{t,s} = & \sum_{i=1}^{nc} \left(\dfrac{p_{i,t,s}^c}{H_i^c \cdot\eta(p_{i,t,s}^c)} + \max(a_i^q \cdot q_{i,t,s}^c - b_i^q,0)\right)\\
        &+ (a^b + b^b \cdot (q_{i,t,s}^b) + c^b (q_{i,t,s}^b))^2) \label{eq: emiobj}
        \end{aligned}
    \end{equation}
The emission objective function can then be written as:
\begin{gather}
    O_2 = \dfrac{1}{|S|} \sum_{s = 1}^S\sum_{t=1}^T \left(g_{t,s} \cdot L_{t,s}  + EF^{grid} \cdot max(p_{t,s}^e,0)\right)
\end{gather}

\textbf{Heat Waste} The heat waste objective is structured to penalize significant levels of waste, allowing a small amount of extra heat that is needed to compensate for transmission loss. Therefore, we only penalize the amount of heat waste when it exceeds the threshold $\delta$, which can be written as:

\begin{gather}
    O_3 = \dfrac{1}{|S|} \sum_{s = 1}^S\sum_{t=1}^T I_{t,s}\\
    I_{t,s} =  \left\{ \begin{array}{rcl}
1, & \mbox{if} \quad 1+\dfrac{\sum_{i=1}^{nc}q_{i,t,s}^c + q_{t,s}^b - HL_{t,s}}{HL_{t,s}} > \delta \\ 
0, & \mbox{otherwise} &
\end{array}\right.  \label{eq: hwobj}
\end{gather}

\subsubsection{System Level Constraints}
In addition to the physical constraints constructed in Section~\ref{sec: Microgridmodel}, two additional constraints are considered as follows.

\textbf{Load Balance} The demand and load have to be balanced for all the time of operation:
    \begin{gather}
        L_{t,s} = \sum_{i=1}^{nc} p_{i,t,s}^c + p_{t,s}^s + y_{t,s}^h +y_{t,s}^s + p_{t,s}^e \label{eq: loadbalance}
    \end{gather}
where $L_{t,s}$ is the electrical demand for hour $t$ in scenario $s$.

\textbf{Heat Reliability}
    To ensure the reliability of the thermal supply, we require that at least 95\% of the heat load has to be satisfied for more than 22 hours of the day. Note that under this constraint, less than 5\% of the load can be shed by lowering building temperature setpoints for a short period of time without indoor temperatures dropping below the desired comfort range.
    \begin{gather}
        \sum_{s = 1}^S\sum_{t=1}^T RL_{t,s} \geq \dfrac{22}{24}\\
        RL_{t,s} = \left\{ \begin{array}{rcl}
            1, & \mbox{if} \quad 1+\dfrac{\sum_{i=1}^{nc}q_{i,t}^c + q_t^b - HL_{t,s}}{HL_{t,s}} \geq 0.95 \\ 
            0, & \mbox{otherwise} &
            \end{array}\right.\label{eq: heatrel}
    \end{gather}

The independent decision variables that agents must decide on are the generation of electricity and steam from CHPs and the generation of steam from the boiler, which can be formulated as $A_t = ({p_{i,t,s}^c,q_{i,t,s}^c,q_{i,t,s}^b})$.  The ST output $p_{t,s}^s$ and the energy exchange $p_{t,s}^e$ depend on the available steam and load as described in Eq.~(\ref{eq: stgen}) and~(\ref{eq: loadbalance}). The multi-objective MG energy management problem can then be formulated as:\\

\begin{equation}
    \begin{aligned}
A_t &= \argmin  \mathbf{F}\\
\mathbf{F} &= (O_1,O_2,O_3)\\
s.t. \quad & (\ref{eq: chpramp}) - (\ref{eq: pcc}),(\ref{eq: loadbalance})-(\ref{eq: heatrel})
\end{aligned} \label{eq: obj+constr}
\end{equation}

\section{Methodologies} \label{sec: method}
In this section, we first introduce the MORL framework supported by the Borg MOEA and reformulate the multi-objective stochastic decision-making problem defined in Eq.~(\ref{eq: obj+constr}) as a model-free RL problem. Then, the technique to perform a time-varying sensitivity analysis that supports the interpretation of the high-dimensional parametric policies derived from the MORL framework is presented.

\subsection{State Variables}
Conventionally, the state of a MDP represents the information or observation that the agent receives when interacting with the environment to make decisions. To deal with the unobservant variables and leverage the exogenous information that might be useful, the state of the system includes three parts: 1) the observable exogenous information $W_t$, 2) the unobservable variables $H_t$, and 3) and the observable action from previous time step $Y_t = A_{t-1}$. Note that only $W_t$ and $Y_t$ are observable to the agent.

For the MG energy management problem defined in this study, the observable exogenous information $W_t$ includes weather forecast and day ahead locational marginal price; $H_t$ consists of uncertainties such as renewable outputs, load, and real-time price signal; and $Y_t$ draws from the actions of time $t-1$, which are part of the state $S_{t}$. The state can then be formulated as:
\begin{equation}
    \begin{aligned}
        S_t &= (W_t,Y_t,H_t)\\
        W_t &= (Te_{t,s},Wv_{t,s},Sr_{t,s},Sf_{t,s},\hat{\pi_{t,s}})\\
        H_t &= (L_{t,s},y_{t,s}^s,y_{t,s}^h,HL_{t,s},\pi_{t,s})\\
        Y_t &= (p_{i,t-1,s}^c,q_{i,t-1,s}^c,q_{i,t-1,s}^b)
    \end{aligned}
\end{equation}
where $Te_{t,s}$, $Wv_{t,s}$,$Sr_{t,s}$,$Sf_{t,s}$ and $\hat{\pi_{t,s}}$ are the temperature, wind speed, solar radiation, stream flow of the local river that provides hydropower and the real-time location marginal price of previous day at hour $t$ for scenario $s$, respectively. 

\subsection{Actions and function approximations} \label{sec: ANN}
Instead of solving for the sequential decisions made for each time step $t$, policy approximation is used here to map the state to the action $A_t = \pi_{\theta}(W_t,Y_t)$. The benefit is three-fold: 1) the function approximation reduces the number of parameters that need to be trained; 2) the trained parametric policy can be generalized to other unobserved system states without repeated training; and 3) the parametric policy can be analyzed to assist interpretation of the behavior of the agent, unveiling the "black-box" for AI methods.

\begin{figure}[h]
  \begin{center}
  \includegraphics[width=0.85\columnwidth]{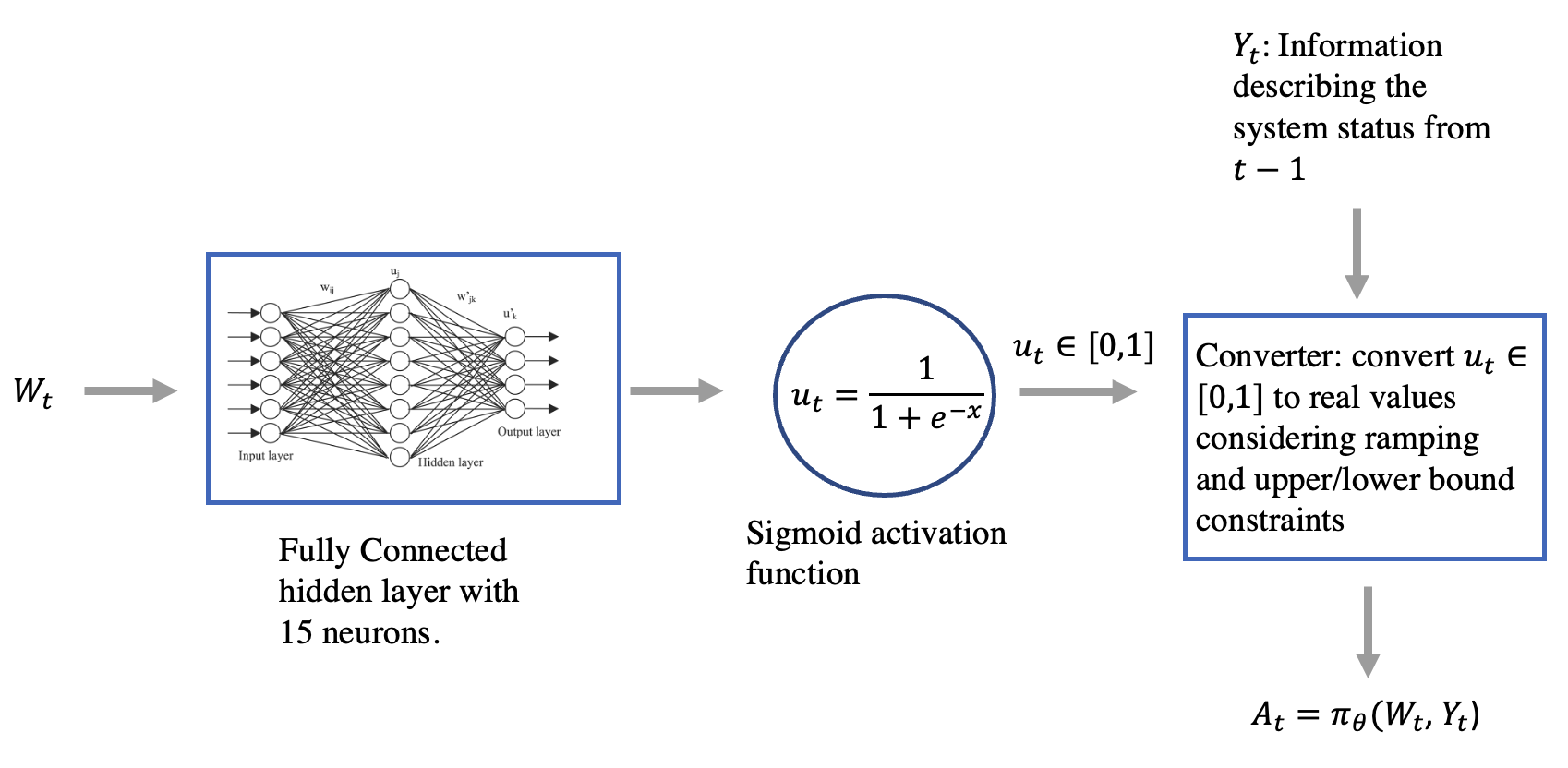}
  \caption{Function approximation for the parametric policy }\label{fig: ANN}
    \vspace{-20pt}
  \end{center}
\end{figure}

In this specific case, a multi-input, multi-output ANN is chosen to coordinate the decisions made by multiple devices in the MG, leveraging the hidden layer to exchange information among the multiple outputs. As shown in Figure~\ref{fig: ANN}, the ANN has one hidden layer with 15 neurons that are fully connected to the input and output layers. The sigmoid function is chosen as the activation function. As a result, the decisions $u_t$ are in the range of [0,1]. A set of weather forecasts are chosen to be the observable part of the state for the system $W_t$ to serve as input for the ANN. The output of the ANN $u_t \in [0,1]$ is then converted to the real value given the ramping and power limit constraints~(\ref{eq: chpramp}), (\ref{eq: chplim}) and~(\ref{eq: boilerlim}) based on $Y_t$. 


\subsection{Reward Functions}
There are three reward functions needed to be evaluated for each state-action pair for the three objective functions formulated in Section~\ref{sec: opffomulation}. The reward functions for hour $t$ in scenario $s$ can be written as:
\begin{equation}
    \begin{aligned}
        R_1(s_t,a_t) &= \sum_{i=1}^{nc} \left(C(p_{i,t,s}^c) +C(q_{i,t,s}^c)\right)+C(q_{t,s}^b)+C(p_{t,s}^e)\\
        R_2(s_t,a_t) &= g_{t,s} \cdot L_{t,s}  + EF^{grid} \cdot max(p_{t,s}^e,0)\\
        R_3(s_t,a_t) &= I_{t,s}
    \end{aligned}
\end{equation}

\subsection{Borg MOEA}
The Borg MOEA~\cite{hadka2013borg} is chosen as the optimization algorithm to handle multiple value functions in place of the conventional gradient-based policy search algorithms to find the best set of weights for the ANN. Similar to other MOEAs, Borg maintains a population of candidate solutions called individuals, which evolves through six operators to generate a diverse set of Pareto-approximate solutions. Compared to other MOEAs that are vulnerable to stalling in local optimal and sensitive to the initialization of hyper-parameters, The Borg MOEA has the advantage of using self-adaptive search operators and $\epsilon$-dominance archiving techniques to overcome these weaknesses~\cite{hadka2012diagnostic}. The Borg MOEA has been used for a wide variety of planning and control applications in water supply and hydropower contexts with strongly complex and non-linear stochastic simulations. It has been carefully benchmarked in its ability to find high-dimensional and geometrically challenging Pareto frontiers~\cite{quinn2017direct,quinn2018exploring}. In this study, we use the multi-master Borg (mm-Borg)~\cite{hadka2015large,giuliani2017scalable}, which is a large-scale parallel implementation of the original Borg algorithm, harnessing the power of computing architectures. Details of the mm-Borg implementation can be found in SI.2.  
\vspace{-5pt}
\subsection{Model-free RL}
\begin{figure}[h]
  \begin{center}
  \includegraphics[width=0.99\columnwidth]{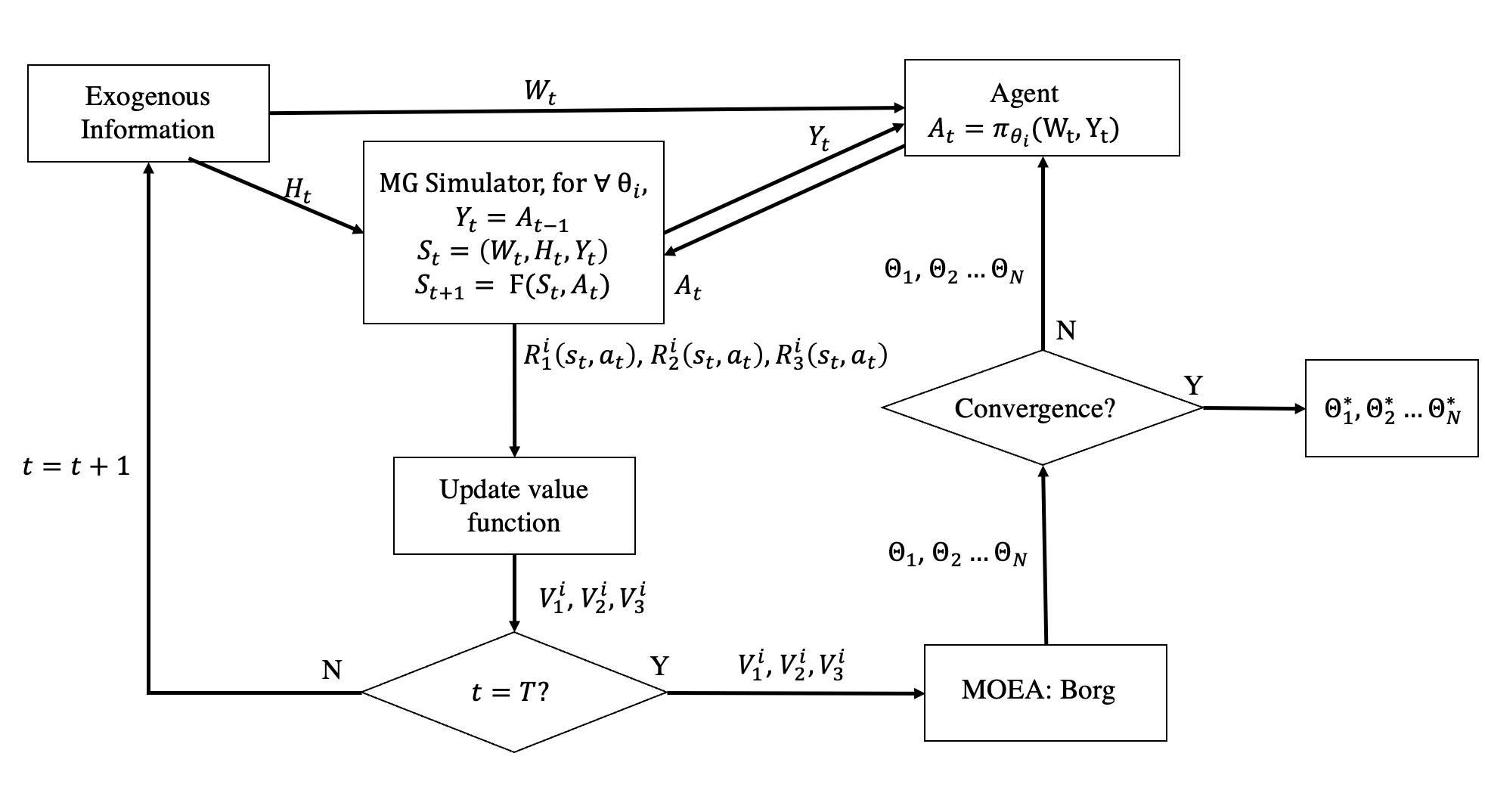}
  \caption{Overview of the MORL framework }\label{fig: MORL}
    \vspace{-20pt}
  \end{center}
\end{figure}

As the underlying model for the state $S_t$ to transit to $S_{t+1}$ is not known, the agent is trained by interacting with the environment to get feedback implicitly from the underlying model $S_{t+1} = f(S_t, A_t)$. As shown in Figure~\ref{fig: MORL}, the agent makes the decision given the current set of parameters $\theta$ and the observable component of the state $W_t$ and  $Y_t = A_{t-1}$. Then the actions $A_t$ are sent to the MG simulator, where the unobservable part of the state $H_t$ interacts with $A_t$ to transit the system state from $S_t$ to $S_{t+1}$ and generates reward $R_1, R_2, R_3$ for time step $t$ given the state-action pair\footnote{Note that the flowchart in Figure~\ref{fig: MORL} is simplified that the scenario loop is not presented. In the computation, the value function is the expectation over all the possible scenarios considered as shown in Eq~\ref{eq: valuefunction}}. The value functions for individual $i$ can be written as the expectation of all the scenarios considered for time horizon $t \in [1, T]$:
\begin{equation}
    \begin{aligned}
        V_1^i = \mathbb{E}_{s}(\sum_{t=1}^{T}R_1(s_t,a_t)) \\
        V_2^i = \mathbb{E}_{s}(\sum_{t=1}^{T}R_2(s_t,a_t)) \\
        V_3^i = \mathbb{E}_{s}(\sum_{t=1}^{T}R_3(s_t,a_t))
    \end{aligned} \label{eq: valuefunction}
\end{equation}

The process is repeated for each scenario $s$ and individual $i$ of the current non-dominated population of size N. The set of non-dominated value functions is sent to the mm-Borg MOEA as feedback to generate the next set of parameters $[\theta_1, \theta_2,..., \theta_N]$ for the ANN. If the algorithm is converged or hits the wall-time, then the set of optimal non-dominated policy parameters $[\theta_1^*, \theta_2^*,..., \theta_N^*]$ is found. 

\subsection{Time Varying Sensitivity Analysis} \label{sec: TVSA}
To investigate how the agent uses the information, i.e. the high-dimensional parametric policies, we perform the Time Varying Sensitivity Analysis (TVSA) which  reveals how the value of exogenous information changes through time. This information is valuable for the local system operator to understand the importance of forecasts and incorporate them into the future decision-making process. By taking advantage of the variance decomposition methods in~\cite{quinn2019controlling}, the conditional variance of the decisions $u_t$ at each time $t$ is decomposed by a Taylor series approximation:
\begin{equation}
    \begin{aligned}
       & Var(u_t|t) = \sum_{a}\Bigl(\dfrac{\partial u_t}{\partial(W_t)_a} \Bigr) Var\bigl((W_t)_a\bigr)\\
        &+\sum_{a}\sum_{b \neq a}\Bigl(\dfrac{\partial u_t}{\partial(W_t)_a} \Bigr)\Bigl(\dfrac{\partial u_t}{\partial(W_t)_b} \Bigr) Cov\bigl((W_t)_a,(W_t)_b\bigr)
    \end{aligned}
\end{equation}
Where the partial derivatives of the policy function, which is the ANN in this study, are taken with respect to each piece of time-varying observable information in $W_t$. Note that the hour of the day is not because it explains zero variables across different days. The conditional variance represents the variance in the decisions explained by the input(s) of interest.

\section{Numerical Results} \label{sec: numericresults}

In this study, the framework described in Section~\ref{sec: method} is applied to the CU-MG to analyze the tradeoffs between the objectives. We demonstrate that the Pareto optimal solutions found by the MORL framework exhibit comprehensive tradeoffs and non-linearities. Representative policies are chosen to show the adaptive and coordinated decisions made by the proposed framework. Additionally, through time-varying sensitivity analysis (TVSA), the temporal dynamics of correlation between inputs and outputs of the policies are established to assist system operators and decision-makers in future MG operations.

\subsection{Training and Test data}
Historical weather data for 2019 from a local weather station (Groton, NY) provides temperature, wind speed, and solar insulation to be used as realistic forecasts to inform agents. Stream flow data is obtained for the Fall Creek River from the US Geological Survey~\cite{stdata} to inform on-site hydro production.  Finally, the NYISO~\cite{nyisodata} real-time locational price from the previous day for the Central Load Zone provides the price forecast. Winter (October to April) and summer (May to September) seasons are considered separately because operational decisions are further complicated by the ability to take generation units offline in the summer season\footnote{Unlike in winter, where both CHP units are required for electricity and steam generation, the lower steam demand in summer makes it desirable to have the option to switch off generators when not needed.}. The data for each season is divided into training and test sets. The training data are used to identify the non-dominated set of parametric policies considering the three potentially conflicting objectives. The test set is subsequently used to validate the generalizability of these policies to unseen data.

\subsection{Tradeoffs in Objective Space}

\begin{figure}[h]
  \begin{center}
  \includegraphics[width=0.75\columnwidth,trim = {3cm 2cm 2cm 2cm},clip]{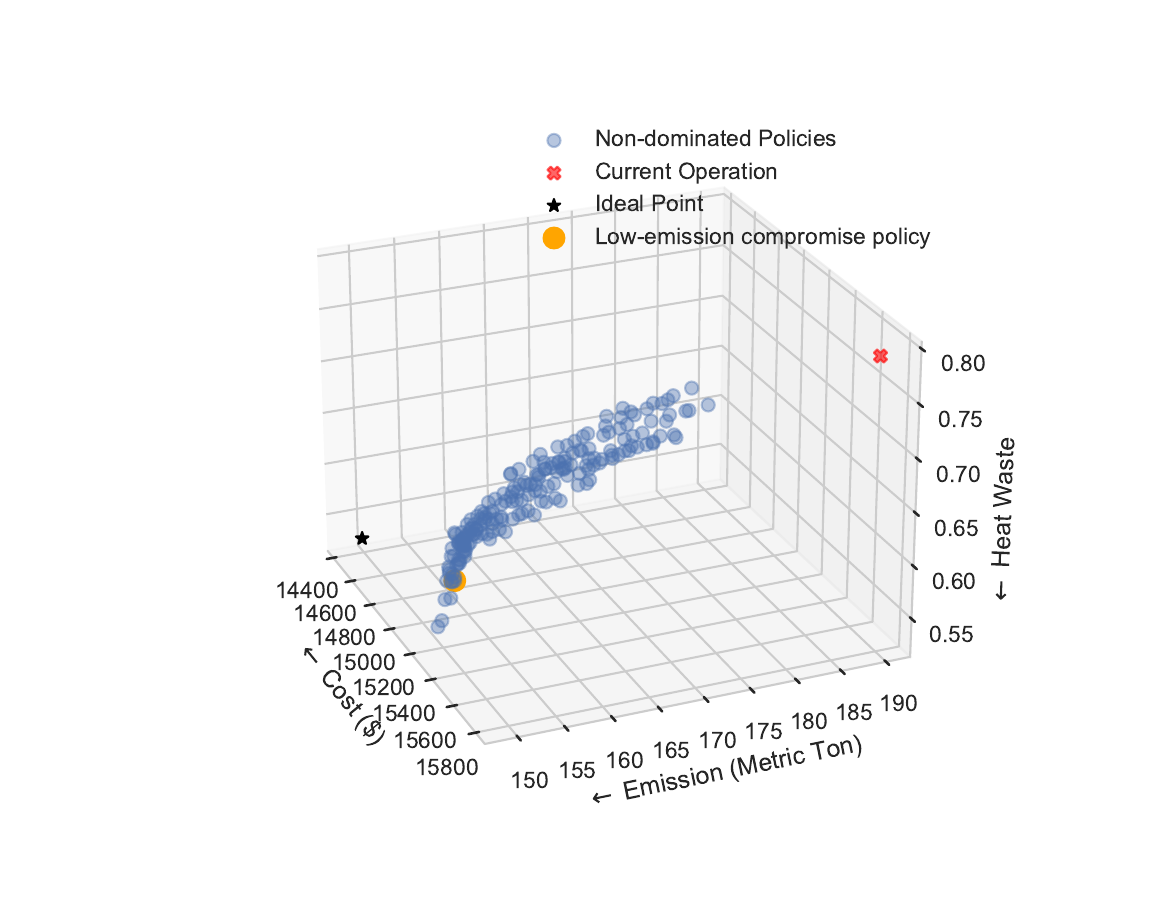}
  \caption{Approximate Pareto Frontier of Winter Results }\label{fig:winterpf}
    \vspace{-10pt}
  \end{center}
\end{figure}

\begin{figure}[h]
  \begin{center}
  \includegraphics[width=0.75\columnwidth,trim = {3cm 1cm 2cm 2cm},clip]{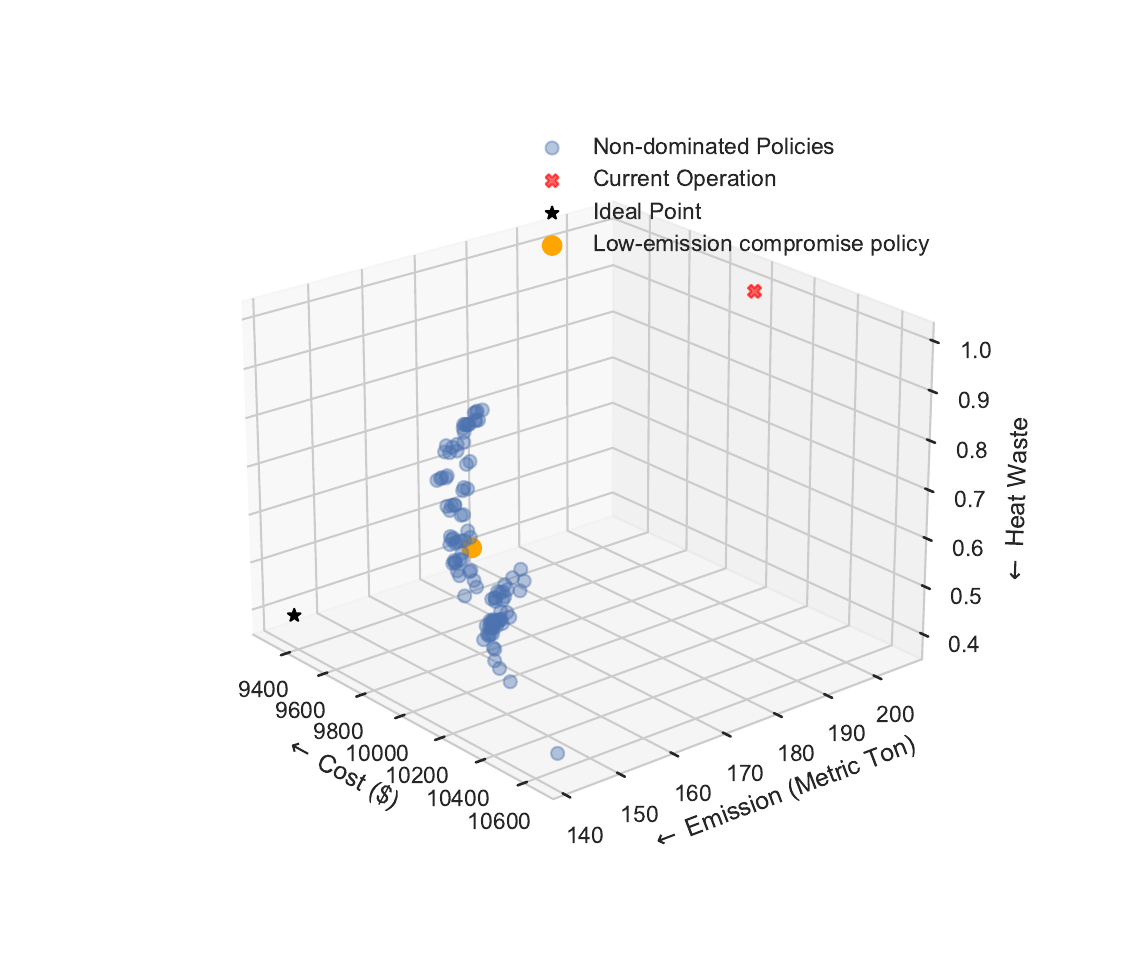}
  \caption{Approximate Pareto Frontier of Summer Results }\label{fig:summerpf}
    \vspace{-10pt}
  \end{center}
\end{figure}

The approximate Pareto frontiers obtained through the MORL framework provided in Figures~\ref{fig:winterpf} and \ref{fig:summerpf} reveal pronounced non-linearity and non-convexity in the tradeoffs among the objectives. In comparison to historical operational strategies, the non-dominated policies derived from the MORL framework consistently exhibit superior performance across all considered objectives. Figure~\ref{fig:winterpf} depicts 186 non-dominated policies for the winter test set, while Figure ~\ref{fig:summerpf} shows 109 non-dominated policies for the summer test set. Each policy is represented by a blue circle representing the average daily behavior of the policy in relation to the three objectives. The black star situated in the bottom left corner denotes the "Ideal Point," which is a hypothetical representation of the theoretical best attainment values across the objectives. Additionally, the red cross in the top right corner represents the average performance of the current operational strategy as observed in historical data.

The summer Pareto Frontier exhibits discontinuity that is not observed in winter, due to the presence of binary decision variables. This characteristic poses challenges to modeling using a weighted sum approach, as discussed in~\cite{salehi2022comprehensive}. The findings indicate that while it is possible to simultaneously reduce emissions, minimize heat waste, and reduce costs from the \emph{current} operational strategy, once operating on the Pareto optimal frontier, achieving a reduction in emissions without increasing costs becomes operationally infeasible. Each of the circles on the Pareto frontier represents an alternative policy, enabling decision-makers to visualize the tradeoffs between objectives and make informed decisions by exploring the full solution space.

To examine the tradeoffs between cost and emission and address the campus decision-makers' interests in emission reduction potential, a representative ``low-emission compromise" policy is selected for each season to demonstrate the potentially achievable performance improvements (highlighted by orange circles in Figures~\ref{fig:winterpf} and~\ref{fig:summerpf}). The comparison between these representative policies and the current operation is presented in Table~\ref{tab: policycomparison}. Without compromising cost, emissions can be reduced by 20\% and 25\% for winter and summer, respectively. It is important to note that these improvements are based solely on the use of weather and historical price information, which are readily available on an hourly basis without the need for generating electricity demand or renewable output predictions. To gain further insights into the behavior of the policies across different scenarios, we next delve into the hourly decisions made by the representative policies in the subsequent subsection.

\begin{table}[h]
\centering
\caption {\label{tab:summercomparison} Comparison between representative policies and the current operation}
\begin{tabular}{lllll}
\hline
\hline
\multirow{2}{*}{Objective} & Cost & Emission& \multirow{2}{*}{Heat Waste} & \multirow{2}{*}{Season}\\   
  & (\$/day) & (MT/day) & & \\   
\hline
\multicolumn{1}{l}{Current operation} & \multicolumn{1}{l}{15,784} & \multicolumn{1}{l}{189.79} & \multicolumn{1}{l}{0.7878} &\multirow{2}{*}{winter}\\
\multicolumn{1}{l}{Representative policy} & \multicolumn{1}{l}{15,201}& \multicolumn{1}{l}{151.36} & \multicolumn{1}{l}{0.5756} \\\hline
\multicolumn{1}{l}{Current operation} &\multicolumn{1}{l}{10,025}  & \multicolumn{1}{l}{205.26} & \multicolumn{1}{l}{0.9923} & \multirow{2}{*}{summer}\\
\multicolumn{1}{l}{Representative policy} &\multicolumn{1}{l}{9,902}  & \multicolumn{1}{l}{154.83} & \multicolumn{1}{l}{0.6083} \\
\hline
   \hline        
        \end{tabular}
    \label{tab: policycomparison}
\end{table}
\vspace{-5pt}

\subsection{Adaptive and Coordinative Performance of Policies} \label{sec: hourlyanalsyis}

\begin{figure*}[h]
  \begin{center}
  \includegraphics[width=0.80\textwidth]{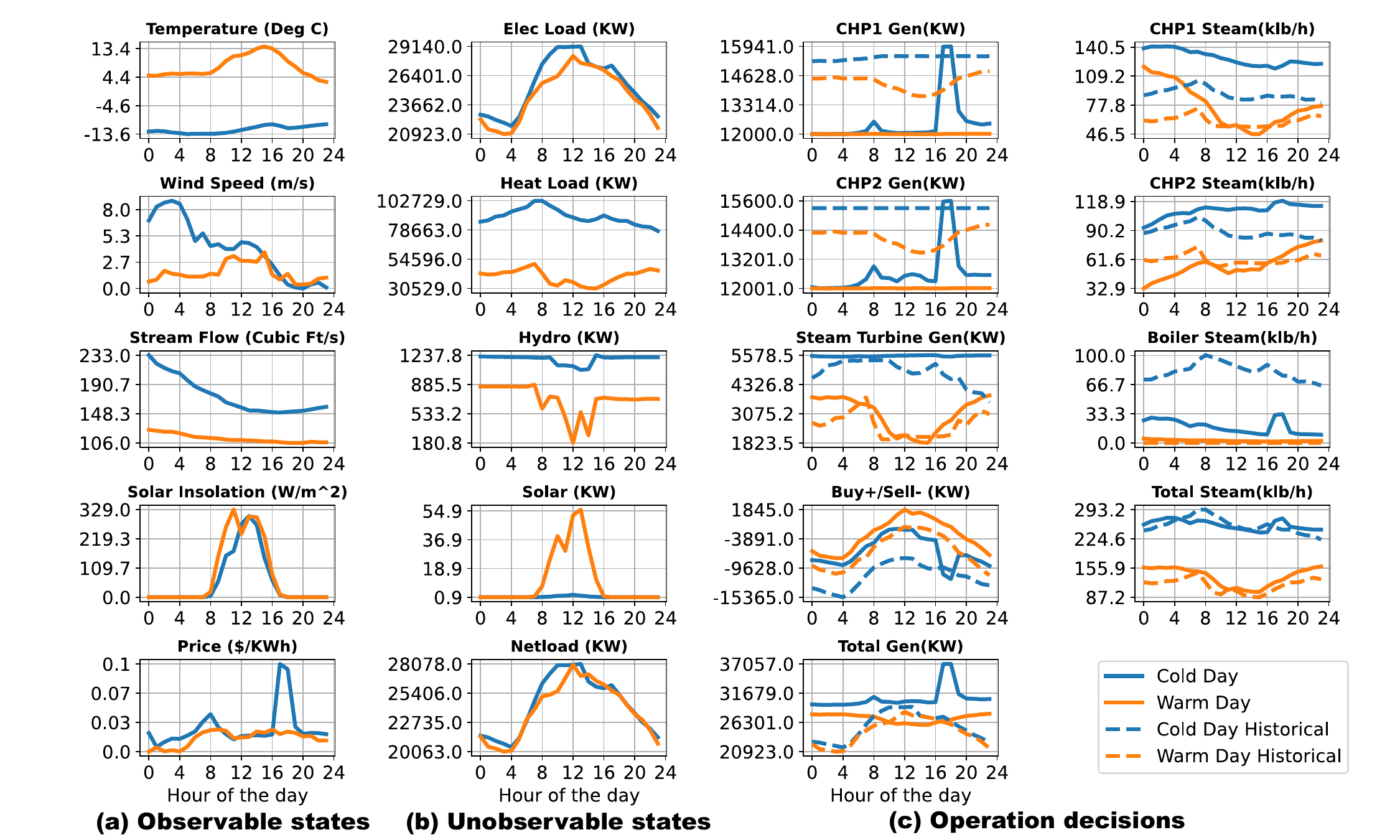}
  \caption{Daily plots for low-emission compromise policy (winter)}\label{fig: winterdaily}
  \vspace{-10pt}
  \end{center}
\end{figure*}

\begin{figure*}[h]
  \begin{center}
  \includegraphics[width=0.8\textwidth]{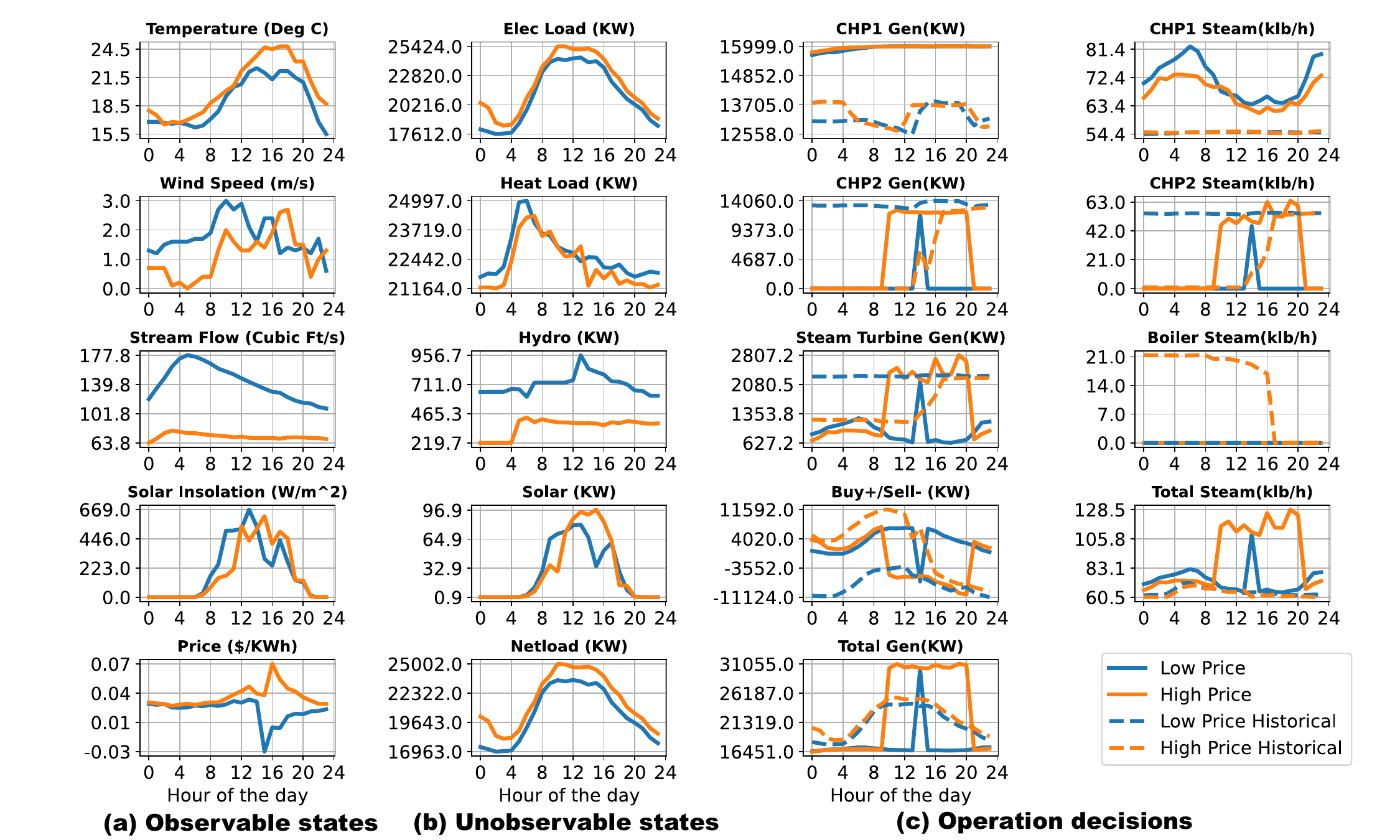}
  \caption{Daily plots for low-emission compromise policy (summer)}\label{fig: summerdaily}
  \vspace{-10pt}
  \end{center}
\end{figure*}

In this section, we consider the detailed hourly decisions made by the representative low-emission compromise policies for both seasons (see Figures 4 and 5). This analysis highlights the adaptive behavior of these policies and demonstrates the use of coordinated decisions among multiple generators to improve overall performance. For a comparative analysis, we select two representative days with distinct temperature profiles for winter and two days with different price signals for summer from the test set. Figures~\ref{fig: winterdaily} and~\ref{fig: summerdaily} present the hourly plots for the winter and summer comparisons, respectively. 

In Figures~\ref{fig: winterdaily} and~\ref{fig: summerdaily}, the first column illustrates the (observable) system states, while the second column shows the (unobservable) exogenous information that interacts with the decisions made by the policy. In the third and fourth columns, the independent decisions made by the agents for power and steam generation, together with the dependent ones, are presented. Solid orange lines indicate warm-day (winter) or high-price-day (summer), while blue lines represent cold-day (winter) or low-price-day (summer). The dashed lines represent historical decisions by CU-MG operators. By examining the hourly behaviors, we make three key observations: 1) adaptivity to exogenous information, 2) coordinated behavior among decisions, and 3) improved performance and efficiency.

We note that policies exhibit effective adaptation to various scenarios in both winter and summer. For example, during a cold day in winter, a price spike is observed between 17:00 and 18:00, and the electricity generation from the CHPs closely follows the price peaks. However, this response is contingent on the cold temperature; conversely, on a warm winter day the CHPs operate at minimum output regardless of price fluctuations. Steam generation, on the other hand, is more responsive to temperature, as we observe lower steam production during the middle of the day when temperature increases and heat load decreases. In the summer case, the introduction of binary variables for the control of the CHP and boiler status is evident in Figure~\ref{fig: summerdaily}. The on-off status is strongly influenced by the price signal. CHP2 remains off for most of the day to minimize emissions and heat waste, but is brought online during price peaks to avoid purchasing additional power from the utility. Despite not having access to the unobservable information (Figure~\ref{fig: summerdaily}(b)), the agent successfully learns the hidden correlation between weather variables and system realizations through interaction with the environment. These results demonstrate that policies provide adaptive decision-making in response to changes in observable system states.

In addition to adaptation, the multiple decisions made by the agent show the ability to coordinate for enhanced overall performance. The architecture of the ANN, as discussed in subsection~\ref{sec: ANN}, enables information sharing among decisions, fostering close collaboration. For example, during a warm day in the winter, the CHPs modulate their steam generation with an increase (decrease) in steam production by CHP1 (CHP2) early in the day, acknowledging that lower steam requirements are met without the boiler during the warmer day. Similar coordinated behavior can be observed in the summer case, reflecting complementary steam generation patterns. This coordination among decisions is key to improving system performance through efficient utilization of resources.

The synthesis of improved performance is achieved through the adaptive and coordinated decision-making process discussed above. Recall that the policies under analysis prioritize reducing emissions. A policy that prioritizes reducing cost has different behavior. For example, CHPs could constantly generate power at a higher level to gain more revenue through energy exports, as shown in a more detailed description in SI.3. Therefore, the agent's decisions aim to minimize overall local generation whenever possible to maintain a low carbon footprint. Note that this approach is based on the assumption that the average emission rate of Zone C in NYS is lower than the on-campus emission rate, which is a valid assumption considering that Zone C has a significant proportion of emission-free generation~\cite{nys2tail}. In both the winter and summer cases, it is economically advantageous to increase power generation during price peaks. The lower-emission policy allows for greater power sales during those periods when profits can be increased without a substantial increase in emissions. In this case, the agent capitalizes on price peaks to offset the higher costs incurred during non-peak hours, which differs from historical operations that consistently favor selling power to the grid during non-peak price hours. In the summer case, the price signal plays a more critical role, particularly when the locational marginal price becomes negative. 
For example, the policy avoids sales to the grid during times of negative LMP, as demonstrated by comparing the blue dashed (historical) and solid (simulated) lines in panel(c) of Figure~\ref{fig: summerdaily}. 

Furthermore, the agent demonstrates a preference for using CHP units to generate steam, particularly during the cold day when the heat load is higher. This decision is driven by the higher efficiency of the additional steam generated by the CHP units (~100\%) instead of the boiler units (~80\%). By prioritizing CHP units, policies are effective in increasing overall system efficiency. 

Lastly, it is important to note that in the summer case, steam generation is substantially higher relative to historical data. This increase in steam generation is a result of maximizing the use of the steam turbine to generate additional electricity. Although this approach may appear inefficient in terms of reducing heat waste, it is a cost-effective strategy in the long run. The policy achieves a lower average accumulated daily heat waste compared to historical operations (as shown in Table~\ref{tab:summercomparison}) by strategically generating high heat waste during specific hours to produce more electricity and offset costs. To gain further insight into the decision-making process of the trained agent over time, in the next section we map the information used by the agent and analyze how, why, and when the various types of observable information are used to make hourly decisions.

\subsection{TVSA for the Parametric Policies}

\begin{figure}[t!]
    \centering
    \begin{subfigure}[t]{0.49\columnwidth}
        \centering
        \includegraphics[width=\textwidth]{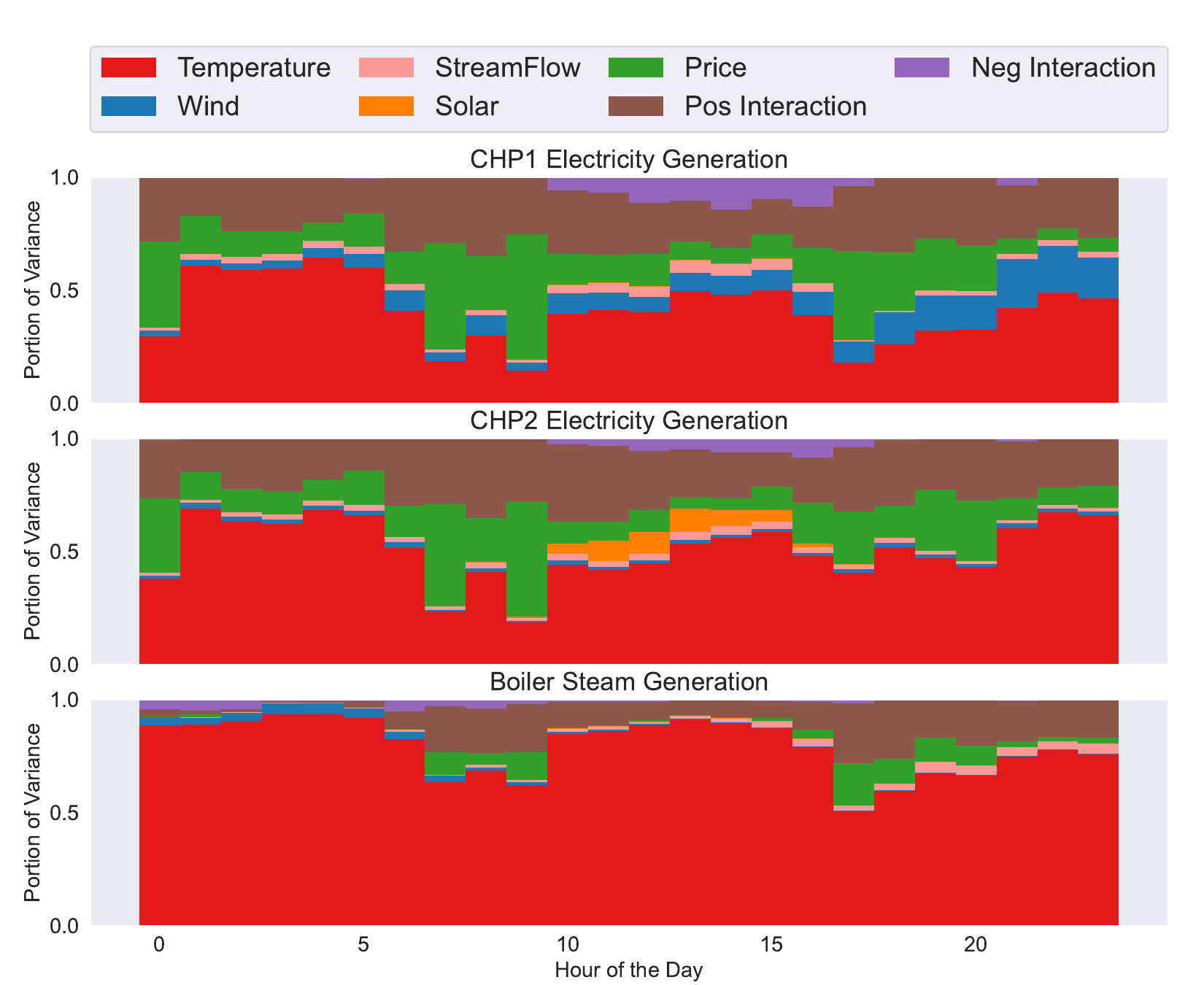}
        \caption{Cold day}
    \end{subfigure}%
    ~ 
    \begin{subfigure}[t]{0.49\columnwidth}
        \centering
        \includegraphics[width=\textwidth]{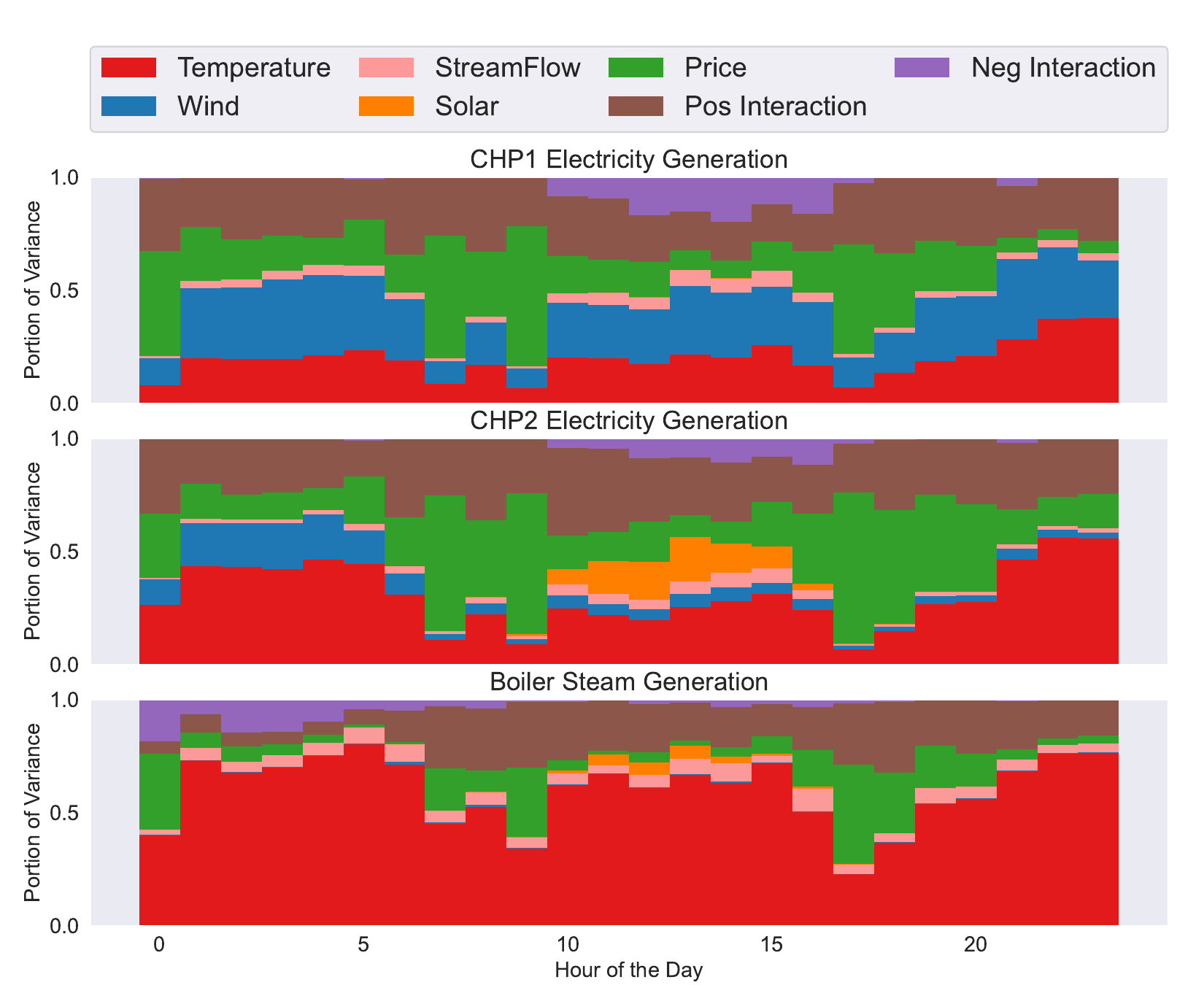}
        \caption{Warm day}
    \end{subfigure}
    \caption{TVSA for winter} \label{fig: TVSAwinter}
      \vspace{-10pt}
\end{figure}

\begin{figure}[h]
  \begin{center}
  \includegraphics[width=0.90\columnwidth]{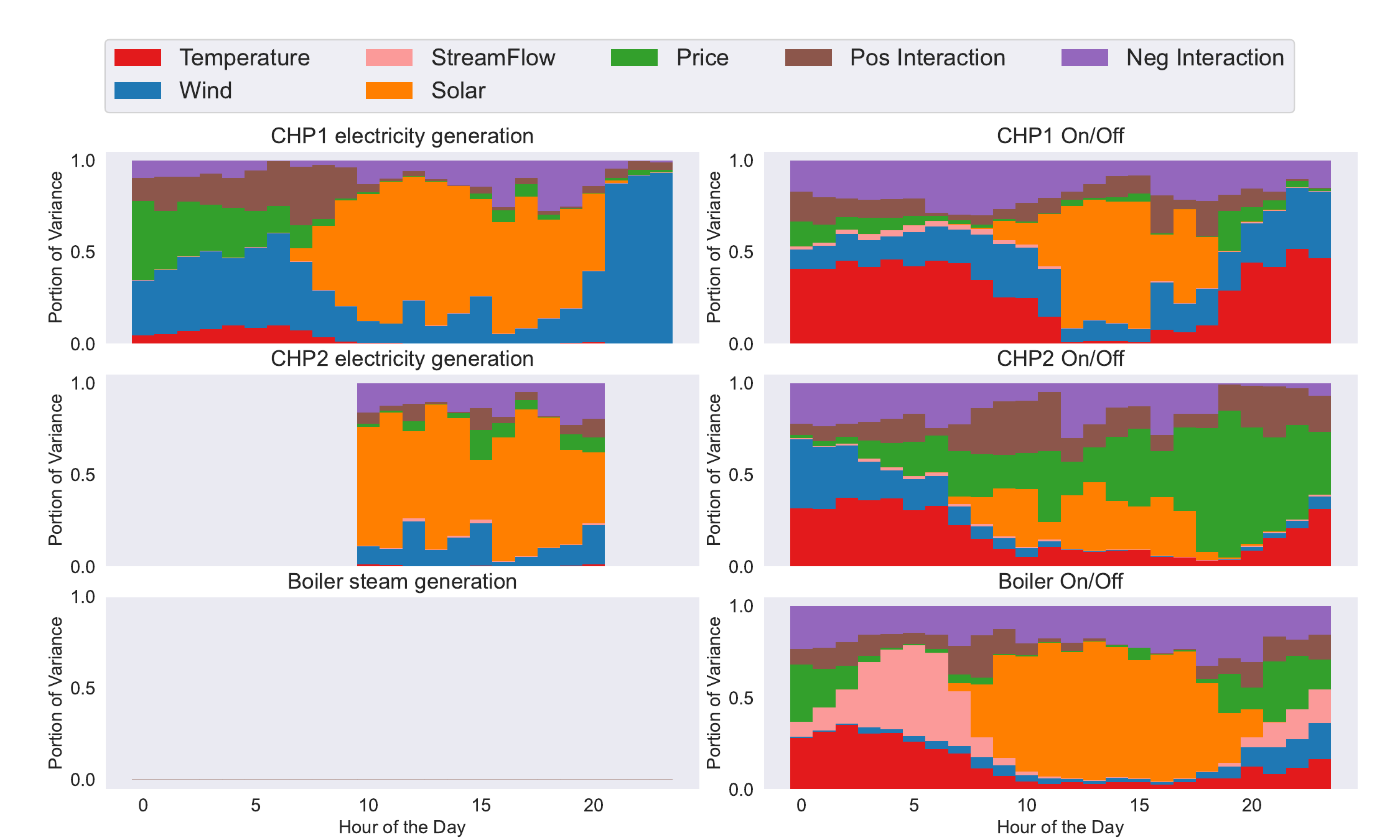}

  \caption{TVSA for high price summer day}\label{fig: TVSAsummerhigh}
  \vspace{-20pt}
  \end{center}
\end{figure}

In this section, we use the Time-Varying Sensitivity Analysis (TVSA) method introduced in Section~\ref{sec: TVSA} to examine the decision-making process and the role of different types of input information. Each stacked bar in Figures~\ref{fig: TVSAwinter} and~\ref{fig: TVSAsummerhigh} presents the normalized conditional variance calculated for input signals, indicating the portion of variance explained by each information signal for selected decisions\footnote{Due to limited space, please refer to figures SI.4 for full representation of all the decisions.}. It is evident that both the price signal and the temperature influence decisions during the winter season. Specifically, temperature has a more pronounced impact during the cold day, supporting our assertion that the agent learns unobservable information (load) based on observable information (temperature). Furthermore, positive interactions between different pairs of information inputs contribute to a considerable amount of explained variance. In particular, the positive interaction is driven primarily by the temperature and price interaction\footnote{Refer to SI.4 for second-order figures.}, which further confirms the findings of Section~\ref{sec: hourlyanalsyis}, highlighting the agent's reliance on conditional information to make informed decisions.

In contrast to winter, summer decisions are driven by a different set of observable information, specifically temperature, solar radiation, and price signal. The introduction of binary variables constrains continuous decisions to zero when the unit is off, resulting in empty cells in the TVSA analysis for CHP2 and the boiler. For binary decisions, solar radiation has a dominant impact, particularly during the midday period. Although solar power constitutes a small portion of the overall electricity supply, the combined influence of solar radiation and temperature shapes the load profile. This observation is further supported by the second-order impact analysis showing that the negative interaction is mostly driven by the temperature and solar radiation pair. The price signal also plays an important role in determining the on-off status of CHP2 throughout the day. Overall, TVSA allows decision-makers to more transparently understand the implications of their solutions of interest as they navigate MG performance tradeoffs. The analysis provides an explicit means for understanding how the dynamics of the different candidate MG control policies' actions are influenced by different exogenous sources of information. The TVSA analysis holds promise to guide the information selection process itself by providing direct insights into the value of candidate exogenous information sources overall.

\section{Conclusion} \label{sec: conclusion}

This paper introduces a MORL framework to tackle the challenges of conflicting objectives in microgrid (MG) energy management problems. The framework leverages the model-free nature of RL and uses observable or easily predictable exogenous information, instead of making predictions for hard-to-predict variables. The value of this exogenous information is further explored through the implementation of TVSA. By harnessing the power of meta-heuristic methods, the proposed approach explicitly searches the entire decision space to identify a more comprehensive set of Pareto-optimal solutions. This enables stakeholders and decision-makers to gain a comprehensive understanding of the tradeoffs between objectives for improved decisions and performance. 

The framework is applied to the CU-MG with the aim of supporting its 2035 carbon-neutrality goal. The results demonstrate that, without the need for additional renewable resources, the proposed framework can achieve a reduction of 20\% and 25\% in carbon emissions during winter and summer, respectively, without incurring additional costs. However, further emission reductions would require careful consideration of the associated cost increase, requiring decision-makers to choose the desired balance from the provided Pareto frontier.

\section*{Acknowledgments}
This work was partially supported by the National Science Foundation Directorate for Computer and Information Science and Engineering Grant ECCS-1453615.


\bibliography{main}

\begin{thebibliography}{10}
\providecommand{\url}[1]{#1}
\csname url@samestyle\endcsname
\providecommand{\newblock}{\relax}
\providecommand{\bibinfo}[2]{#2}
\providecommand{\BIBentrySTDinterwordspacing}{\spaceskip=0pt\relax}
\providecommand{\BIBentryALTinterwordstretchfactor}{4}
\providecommand{\BIBentryALTinterwordspacing}{\spaceskip=\fontdimen2\font plus
\BIBentryALTinterwordstretchfactor\fontdimen3\font minus
  \fontdimen4\font\relax}
\providecommand{\BIBforeignlanguage}[2]{{%
\expandafter\ifx\csname l@#1\endcsname\relax
\typeout{** WARNING: IEEEtran.bst: No hyphenation pattern has been}%
\typeout{** loaded for the language `#1'. Using the pattern for}%
\typeout{** the default language instead.}%
\else
\language=\csname l@#1\endcsname
\fi
#2}}
\providecommand{\BIBdecl}{\relax}
\BIBdecl

\bibitem{lasseter2002microgrids}
R.~H. Lasseter, ``Microgrids,'' in \emph{2002 IEEE power engineering society
  winter meeting. Conference proceedings (Cat. No. 02CH37309)}, vol.~1.\hskip
  1em plus 0.5em minus 0.4em\relax IEEE, 2002, pp. 305--308.

\bibitem{majzoobi2016application}
A.~Majzoobi and A.~Khodaei, ``Application of microgrids in supporting
  distribution grid flexibility,'' \emph{IEEE Transactions on Power Systems},
  vol.~32, no.~5, pp. 3660--3669, 2016.

\bibitem{zhou2015overview}
X.~Zhou, T.~Guo, and Y.~Ma, ``An overview on microgrid technology,'' in
  \emph{2015 IEEE international conference on mechatronics and automation
  (ICMA)}.\hskip 1em plus 0.5em minus 0.4em\relax IEEE, 2015, pp. 76--81.

\bibitem{stvrelec2012modeling}
M.~St{\v{r}}elec, K.~Macek, and A.~Abate, ``Modeling and simulation of a
  microgrid as a stochastic hybrid system,'' in \emph{2012 3rd IEEE PES
  Innovative Smart Grid Technologies Europe (ISGT Europe)}.\hskip 1em plus
  0.5em minus 0.4em\relax IEEE, 2012, pp. 1--9.

\bibitem{wu2013coordinated}
J.~Wu and X.~Guan, ``Coordinated multi-microgrids optimal control algorithm for
  smart distribution management system,'' \emph{IEEE Transactions on Smart
  Grid}, vol.~4, no.~4, pp. 2174--2181, 2013.

\bibitem{puterman1990markov}
M.~L. Puterman, ``Markov decision processes,'' \emph{Handbooks in operations
  research and management science}, vol.~2, pp. 331--434, 1990.

\bibitem{zeng2018dynamic}
P.~Zeng, H.~Li, H.~He, and S.~Li, ``Dynamic energy management of a microgrid
  using approximate dynamic programming and deep recurrent neural network
  learning,'' \emph{IEEE Transactions on Smart Grid}, vol.~10, no.~4, pp.
  4435--4445, 2018.

\bibitem{zhang2019deep}
Z.~Zhang, D.~Zhang, and R.~C. Qiu, ``Deep reinforcement learning for power
  system applications: An overview,'' \emph{CSEE Journal of Power and Energy
  Systems}, vol.~6, no.~1, pp. 213--225, 2019.

\bibitem{mousavi2018deep}
S.~S. Mousavi, M.~Schukat, and E.~Howley, ``Deep reinforcement learning: an
  overview,'' in \emph{Proceedings of SAI Intelligent Systems Conference
  (IntelliSys) 2016: Volume 2}.\hskip 1em plus 0.5em minus 0.4em\relax
  Springer, 2018, pp. 426--440.

\bibitem{zhang2019learning}
Q.~Zhang, K.~Dehghanpour, Z.~Wang, and Q.~Huang, ``A learning-based power
  management method for networked microgrids under incomplete information,''
  \emph{IEEE Transactions on Smart Grid}, vol.~11, no.~2, pp. 1193--1204, 2019.

\bibitem{liu2018distributed}
W.~Liu, P.~Zhuang, H.~Liang, J.~Peng, and Z.~Huang, ``Distributed economic
  dispatch in microgrids based on cooperative reinforcement learning,''
  \emph{IEEE transactions on neural networks and learning systems}, vol.~29,
  no.~6, pp. 2192--2203, 2018.

\bibitem{chen2022reinforcement}
X.~Chen, G.~Qu, Y.~Tang, S.~Low, and N.~Li, ``Reinforcement learning for
  selective key applications in power systems: Recent advances and future
  challenges,'' \emph{IEEE Transactions on Smart Grid}, 2022.

\bibitem{powell2019unified}
W.~B. Powell, ``A unified framework for stochastic optimization,''
  \emph{European Journal of Operational Research}, vol. 275, no.~3, pp.
  795--821, 2019.

\bibitem{bertsekas2019reinforcement}
D.~Bertsekas, \emph{Reinforcement learning and optimal control}.\hskip 1em plus
  0.5em minus 0.4em\relax Athena Scientific, 2019.

\bibitem{khan2016compendium}
A.~A. Khan, M.~Naeem, M.~Iqbal, S.~Qaisar, and A.~Anpalagan, ``A compendium of
  optimization objectives, constraints, tools and algorithms for energy
  management in microgrids,'' \emph{Renewable and Sustainable Energy Reviews},
  vol.~58, pp. 1664--1683, 2016.

\bibitem{woodruff2013many}
M.~J. Woodruff, P.~M. Reed, and T.~W. Simpson, ``Many objective visual
  analytics: rethinking the design of complex engineered systems,''
  \emph{Structural and Multidisciplinary Optimization}, vol.~48, pp. 201--219,
  2013.

\bibitem{gupta2020exploring}
A.~Gupta, M.~Liu, D.~Gold, P.~Reed, and C.~L. Anderson, ``Exploring a direct
  policy search framework for multiobjective optimization of a microgrid energy
  management system,'' 2020.

\bibitem{castelletti2013multiobjective}
A.~Castelletti, F.~Pianosi, and M.~Restelli, ``A multiobjective reinforcement
  learning approach to water resources systems operation: Pareto frontier
  approximation in a single run,'' \emph{Water Resources Research}, vol.~49,
  no.~6, pp. 3476--3486, 2013.

\bibitem{giuliani2021state}
M.~Giuliani, J.~Lamontagne, P.~Reed, and A.~Castelletti, ``A state-of-the-art
  review of optimal reservoir control for managing conflicting demands in a
  changing world,'' \emph{Water Resources Research}, vol.~57, no.~12, p.
  e2021WR029927, 2021.

\bibitem{yang2021multi}
X.~Yang, Z.~Leng, S.~Xu, C.~Yang, L.~Yang, K.~Liu, Y.~Song, and L.~Zhang,
  ``Multi-objective optimal scheduling for cchp microgrids considering
  peak-load reduction by augmented $\varepsilon$-constraint method,''
  \emph{Renewable Energy}, vol. 172, pp. 408--423, 2021.

\bibitem{pourghasem2019stochastic}
P.~Pourghasem, F.~Sohrabi, M.~Abapour, and B.~Mohammadi-Ivatloo, ``Stochastic
  multi-objective dynamic dispatch of renewable and chp-based islanded
  microgrids,'' \emph{Electric Power Systems Research}, vol. 173, pp. 193--201,
  2019.

\bibitem{saberi2019optimal}
K.~Saberi, H.~Pashaei-Didani, R.~Nourollahi, K.~Zare, and S.~Nojavan, ``Optimal
  performance of cchp based microgrid considering environmental issue in the
  presence of real time demand response,'' \emph{Sustainable cities and
  society}, vol.~45, pp. 596--606, 2019.

\bibitem{conti2012optimal}
S.~Conti, R.~Nicolosi, S.~Rizzo, and H.~Zeineldin, ``Optimal dispatching of
  distributed generators and storage systems for mv islanded microgrids,''
  \emph{IEEE Transactions on Power Delivery}, vol.~27, no.~3, pp. 1243--1251,
  2012.

\bibitem{salehi2022comprehensive}
N.~Salehi, H.~Mart{\'\i}nez-Garc{\'\i}a, G.~Velasco-Quesada, and J.~M.
  Guerrero, ``A comprehensive review of control strategies and optimization
  methods for individual and community microgrids,'' \emph{IEEE access},
  vol.~10, pp. 15\,935--15\,955, 2022.

\bibitem{deb2002fast}
K.~Deb, A.~Pratap, S.~Agarwal, and T.~Meyarivan, ``A fast and elitist
  multiobjective genetic algorithm: Nsga-ii,'' \emph{IEEE transactions on
  evolutionary computation}, vol.~6, no.~2, pp. 182--197, 2002.

\bibitem{coello2002mopso}
C.~C. Coello and M.~S. Lechuga, ``Mopso: A proposal for multiple objective
  particle swarm optimization,'' in \emph{Proceedings of the 2002 Congress on
  Evolutionary Computation. CEC'02 (Cat. No. 02TH8600)}, vol.~2.\hskip 1em plus
  0.5em minus 0.4em\relax IEEE, 2002, pp. 1051--1056.

\bibitem{teo2020optimization}
T.~T. Teo, T.~Logenthiran, W.~L. Woo, K.~Abidi, T.~John, N.~S. Wade, D.~M.
  Greenwood, C.~Patsios, and P.~C. Taylor, ``Optimization of fuzzy
  energy-management system for grid-connected microgrid using nsga-ii,''
  \emph{IEEE transactions on cybernetics}, vol.~51, no.~11, pp. 5375--5386,
  2020.

\bibitem{elgammal2018energy}
A.~Elgammal and M.~El-Naggar, ``Energy management in smart grids for the
  integration of hybrid wind--pv--fc--battery renewable energy resources using
  multi-objective particle swarm optimisation (mopso),'' \emph{The Journal of
  Engineering}, vol. 2018, no.~11, pp. 1806--1816, 2018.

\bibitem{WEITZEL2018582}
\BIBentryALTinterwordspacing
T.~Weitzel and C.~H. Glock, ``Energy management for stationary electric energy
  storage systems: A systematic literature review,'' \emph{European Journal of
  Operational Research}, vol. 264, no.~2, pp. 582--606, 2018. [Online].
  Available:
  \url{https://www.sciencedirect.com/science/article/pii/S0377221717305933}
\BIBentrySTDinterwordspacing

\bibitem{hadka2013borg}
D.~Hadka and P.~Reed, ``Borg: An auto-adaptive many-objective evolutionary
  computing framework,'' \emph{Evolutionary computation}, vol.~21, no.~2, pp.
  231--259, 2013.

\bibitem{reed2013evolutionary}
P.~M. Reed, D.~Hadka, J.~D. Herman, J.~R. Kasprzyk, and J.~B. Kollat,
  ``Evolutionary multiobjective optimization in water resources: The past,
  present, and future,'' \emph{Advances in water resources}, vol.~51, pp.
  438--456, 2013.

\bibitem{gupta2020can}
R.~S. Gupta, A.~L. Hamilton, P.~M. Reed, and G.~W. Characklis, ``Can modern
  multi-objective evolutionary algorithms discover high-dimensional financial
  risk portfolio tradeoffs for snow-dominated water-energy systems?''
  \emph{Advances in water resources}, vol. 145, p. 103718, 2020.

\bibitem{salazar2016diagnostic}
J.~Z. Salazar, P.~M. Reed, J.~D. Herman, M.~Giuliani, and A.~Castelletti, ``A
  diagnostic assessment of evolutionary algorithms for multi-objective surface
  water reservoir control,'' \emph{Advances in water resources}, vol.~92, pp.
  172--185, 2016.

\bibitem{beckers2015hybrid}
K.~F. Beckers, M.~Z. Lukawski, G.~A. Aguirre, S.~D. Hillson, and J.~W. Tester,
  ``Hybrid low-grade geothermal-biomass systems for direct-use and
  co-generation: from campus demonstration to nationwide energy player,'' in
  \emph{Proceedings, 40th workshop on geothermal reservoir engineering,
  Stanford University, Stanford, California}, 2015.

\bibitem{aluisio2017optimization}
B.~Aluisio, M.~Dicorato, G.~Forte, and M.~Trovato, ``An optimization procedure
  for microgrid day-ahead operation in the presence of chp facilities,''
  \emph{Sustainable Energy, Grids and Networks}, vol.~11, pp. 34--45, 2017.

\bibitem{hadka2012diagnostic}
D.~Hadka and P.~Reed, ``Diagnostic assessment of search controls and failure
  modes in many-objective evolutionary optimization,'' \emph{Evolutionary
  computation}, vol.~20, no.~3, pp. 423--452, 2012.

\bibitem{quinn2017direct}
J.~D. Quinn, P.~M. Reed, and K.~Keller, ``Direct policy search for robust
  multi-objective management of deeply uncertain socio-ecological tipping
  points,'' \emph{Environmental Modelling \& Software}, vol.~92, pp. 125--141,
  2017.

\bibitem{quinn2018exploring}
J.~D. Quinn, P.~M. Reed, M.~Giuliani, A.~Castelletti, J.~W. Oyler, and R.~E.
  Nicholas, ``Exploring how changing monsoonal dynamics and human pressures
  challenge multireservoir management for flood protection, hydropower
  production, and agricultural water supply,'' \emph{Water Resources Research},
  vol.~54, no.~7, pp. 4638--4662, 2018.

\bibitem{hadka2015large}
D.~Hadka and P.~Reed, ``Large-scale parallelization of the borg multiobjective
  evolutionary algorithm to enhance the management of complex environmental
  systems,'' \emph{Environmental Modelling \& Software}, vol.~69, pp. 353--369,
  2015.

\bibitem{giuliani2017scalable}
M.~Giuliani, J.~D. Quinn, J.~D. Herman, A.~Castelletti, and P.~M. Reed,
  ``Scalable multiobjective control for large-scale water resources systems
  under uncertainty,'' \emph{IEEE Transactions on Control Systems Technology},
  vol.~26, no.~4, pp. 1492--1499, 2017.

\bibitem{quinn2019controlling}
J.~D. Quinn, P.~M. Reed, M.~Giuliani, and A.~Castelletti, ``What is controlling
  our control rules? opening the black box of multireservoir operating policies
  using time-varying sensitivity analysis,'' \emph{Water Resources Research},
  vol.~55, no.~7, pp. 5962--5984, 2019.

\bibitem{stdata}
{U.S. Geological Survey}, ``{Fall Creek Near Ithaca NY - 04234000},''
  \url{https://waterdata.usgs.gov/monitoring-location/03351500/#parameterCode=00065&period=P7D},
  accessed: 2021-12-28.

\bibitem{nyisodata}
{New York Independent System Operator}, ``{New York ISO | Energy and Market
  Operational Data},''
  \url{https://www.nyiso.com/energy-market-operational-data}, accessed:
  2021-06-28.

\bibitem{nys2tail}
------, ``New transmission investments add a new chapter to the “tale of two
  grids”,''
  \url{https://www.nyiso.com/-/new-transmission-investments-add-a-new-chapter-to-the-tale-of-two-grids-},
  2022, accessed: 2023-05-23.

\end{thebibliography}


\begin{thebibliography}{1}
\newcommand{\enquote}[1]{``#1''}

\bibitem{mgdata}
{Energy Management and Control System Cornell}, \enquote{{Energy Management and
  Control System Portal},}
  \url{https://portal.emcs.cornell.edu/d/2/dashboard-list?orgId=2}. Accessed:
  2021-11-28.

\bibitem{holland1977robust}
P.~W. Holland and R.~E. Welsch, \enquote{Robust regression using iteratively
  reweighted least-squares,} {\protect\JournalTitle{Communications in
  Statistics-theory and Methods}} \textbf{6}, 813--827 (1977).

\bibitem{dumouchel1989integrating}
W.~Dumouchel, F.~O’brien \emph{et~al.}, \enquote{Integrating a robust option
  into a multiple regression computing environment,} in \emph{Computer science
  and statistics: Proceedings of the 21st symposium on the interface,}
  (American Statistical Association Alexandria, 1989), pp. 297--302.

\bibitem{ngemrate}
{U.S. Energy Information Administration}, \enquote{{Carbon Dioxide Emissions
  Coefficients},}
  \url{https://www.eia.gov/environment/emissions/co2_vol_mass.php}. Accessed:
  2021-06-28.

\end{thebibliography}
\bibliographystyle{IEEEtran}

\newpage

 




\vfill

\end{document}


\maketitle

\section{Microgrid parameters and generator modeling}
The electricity efficiency, fuel function for extra steam generation, heat rate for the boiler and the electricity recovery function for the steam turbines are estimated using historical data of 2019 from the Cornell microgrid data archive~\cite{mgdata}. The parameterized functions are described in Section II.A in the manuscript and are fitted using the Fit nonlinear regression model (fitnlm) function in Matlab, which is solved by the Iteratively Reweighted Least Squares method~\cite{holland1977robust, dumouchel1989integrating}. The estimated parameters are summarized in Table~\ref{tab:fittedparam}. 

\begin{table}[H]
\centering
\caption {\label{tab:fittedparam} Fitted parameters for generators}

\begin{tabular}{ll}
\hline
\hline
Parameter & value \\   \hline
$a_1^c$  & 0.088094\\
$b_1^c$  & 0.42435\\
$c_1^c$  & 0.19291\\
$a_2^c$  & -0.027957\\
$b_2^c$  & 0.80107\\
$c_2^c$  & 0.34667\\
$a_1^q$  & 1.1766\\
$b_1^q$  & 65.881\\
$a_2^q$  & 1.3293\\
$b_2^q$  & 77.25\\
$a_1$  & 0.0009\\
$b_1$  & 1.0968\\
$c_1$  & 3.7742\\
$a_1^s$  & -1.9341\\
$b_1^s$  & 6042.6\\
$a_2^s$  & 33.907\\
$b_2^s$  & 1552.2\\
$c^s$  & 215\\
\hline
\hline
\end{tabular}
\end{table}

Other microgrid-related parameters are reported in Table~\ref{tab: mgparam} for reproducibility. 
\begin{table}[H]
\centering
\caption {\label{tab: mgparam} Microgrid related parameters}

\begin{tabular}{ll}
\hline
\hline
Parameter & value \\   \hline
$\underline{P_i^c}$  & 12000 KW\\
$\overline{P_i^c}$  & 16000 KW\\
$\underline{R_i^c}$  & -5000 KW\\
$\overline{R_i^c}$  & 5000 KW\\

$\underline{Q_i^c}$  & 0 klb/h \\
$\overline{Q_i^c}$  & 153 klb/h\\

$\underline{Q^b}$  & 0 klb/h\\
$\overline{Q^b}$  & 540 klb/h\\

$H_c$  & 293 kwh/dth\\
$\epsilon$  & 116.65 lb/dth~\cite{ngemrate} \\
$\delta$  & 1.05\\
$EF^{grid}$  & 0.932 lb/kwh\\

\hline
\hline
\end{tabular}
\end{table}

\section{Parameters for mm-Borg Implementation}
The mm-Borg is implemented with four masters to run in parallel to find the best Pareto optimal solution set. The epsilons for cost, emission, and heat waste are set to be 10, 1, and 0.01 to filter out the $\epsilon$-dominate solutions. As the MOEAs are sensitive to the initialization of hyperparameters, 10 random seeds are used to initialize 10 parallel runs, each with 500,000 functions of evaluations. The final result is the joint Pareto frontier over the solutions from all the random seeds.

\section{High-emission, Low-cost policy analysis}
\begin{figure}[H]
  \begin{center}
  \includegraphics[width=0.99\textwidth]{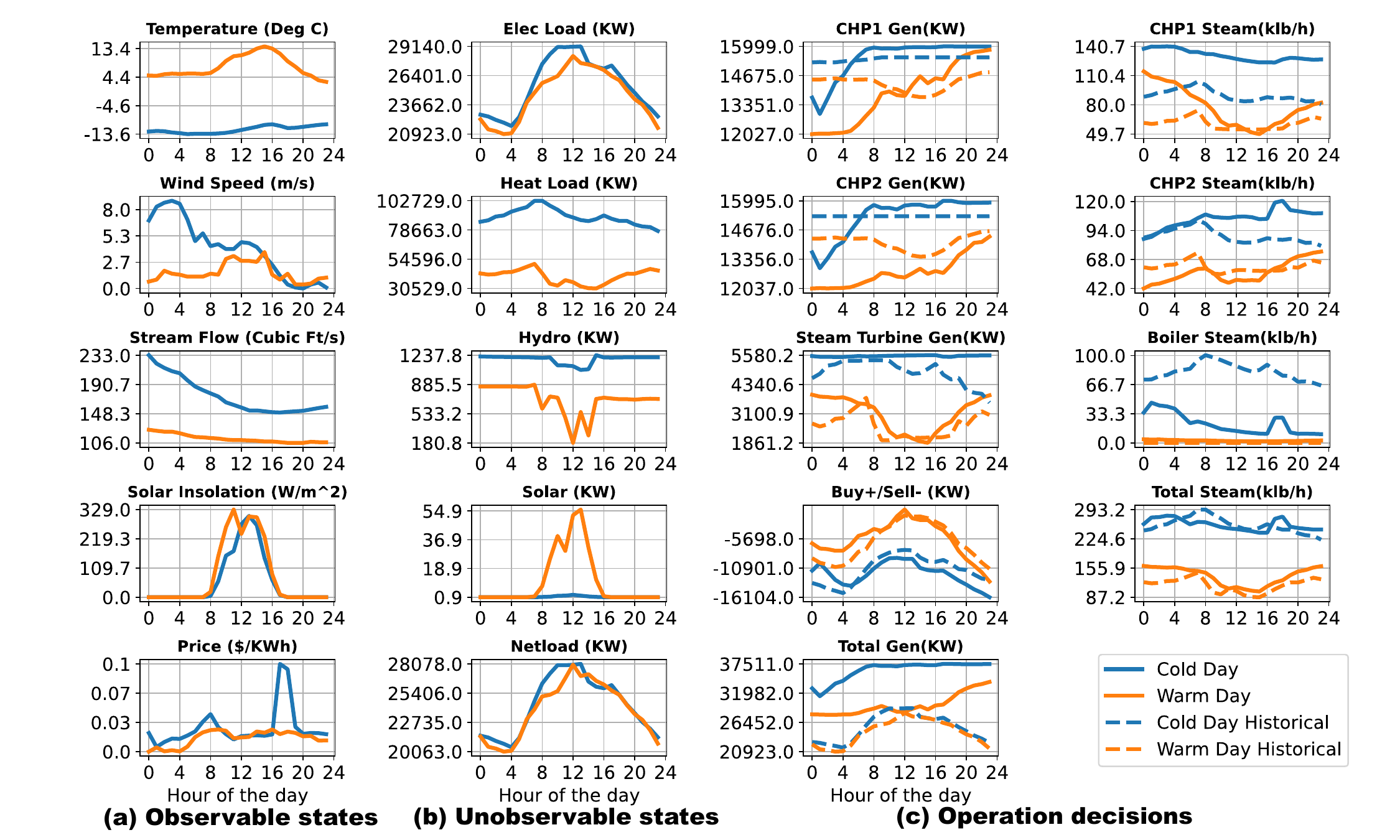}
  \caption{Daily plots for low cost high emission policy in winter}\label{fig: winterhighemission}
  \vspace{-10pt}
  \end{center}
\end{figure}

To analyze the behavior of a policy with different preferences, we selected a low-cost, high-emission policy and compared its average daily performance across three objectives with historical operation, as summarized in Table~\ref{tab: winterhighemission}. Remarkably, an approximately 8\% reduction in cost was achieved without compromising emissions. The corresponding hourly decisions for both cold and warm days are illustrated in Figure~\ref{fig: winterhighemission}. It is evident that both CHPs are generating more electricity locally, enabling a higher volume of power to be sold to the utility rather than purchased. This Buy/Sell behavior closely aligns with the current operation, compared to the high-cost, low-emission policy. This is as expected because the current operation targets minimizing cost only. The cost reduction is primarily attained by generating slightly more power, particularly during price peaks. However, as the policy's priority remains high-level generation most of the time, electricity generation closely follows the electricity load curve to minimize emissions, especially during periods of low prices. Recall that the load profile is not available to the agent until it makes the decisions, which means the agent learned the latent load profile by using the observable weather and price variables. Steam generation exhibits similarities to the high-cost, low-emission policy, with the coordinated operation of the CHPs and the boiler aligning with the heat load while prioritizing the utilization of CHPs that have higher efficiency. .

\begin{table}[h]
\centering
\caption {\label{tab: winterhighemission} Winter Compassion between representative policies and the current Operation}
\resizebox{\columnwidth}{!}{
\begin{tabular}{llll}
\hline
\hline
Objective & Cost (\$/day) & Emission (MT/day) & Heat Waste\\   \hline
 Current Operation & 15,784 & 189.79 & 0.7878\\
Low cost, high emission & 14,462 & 182.14 & 0.5782\\
\hline
\hline
\end{tabular}}
\end{table}

\section{Second-order TVSA analysis} 
This note summarizes first- and second-order TVSA for the warm and cold days in winter and the high-price and low-price days in summer under the high-cost, low-emission policies. 

\begin{figure}[H]
  \begin{center}
  \includegraphics[width=0.9\textwidth]{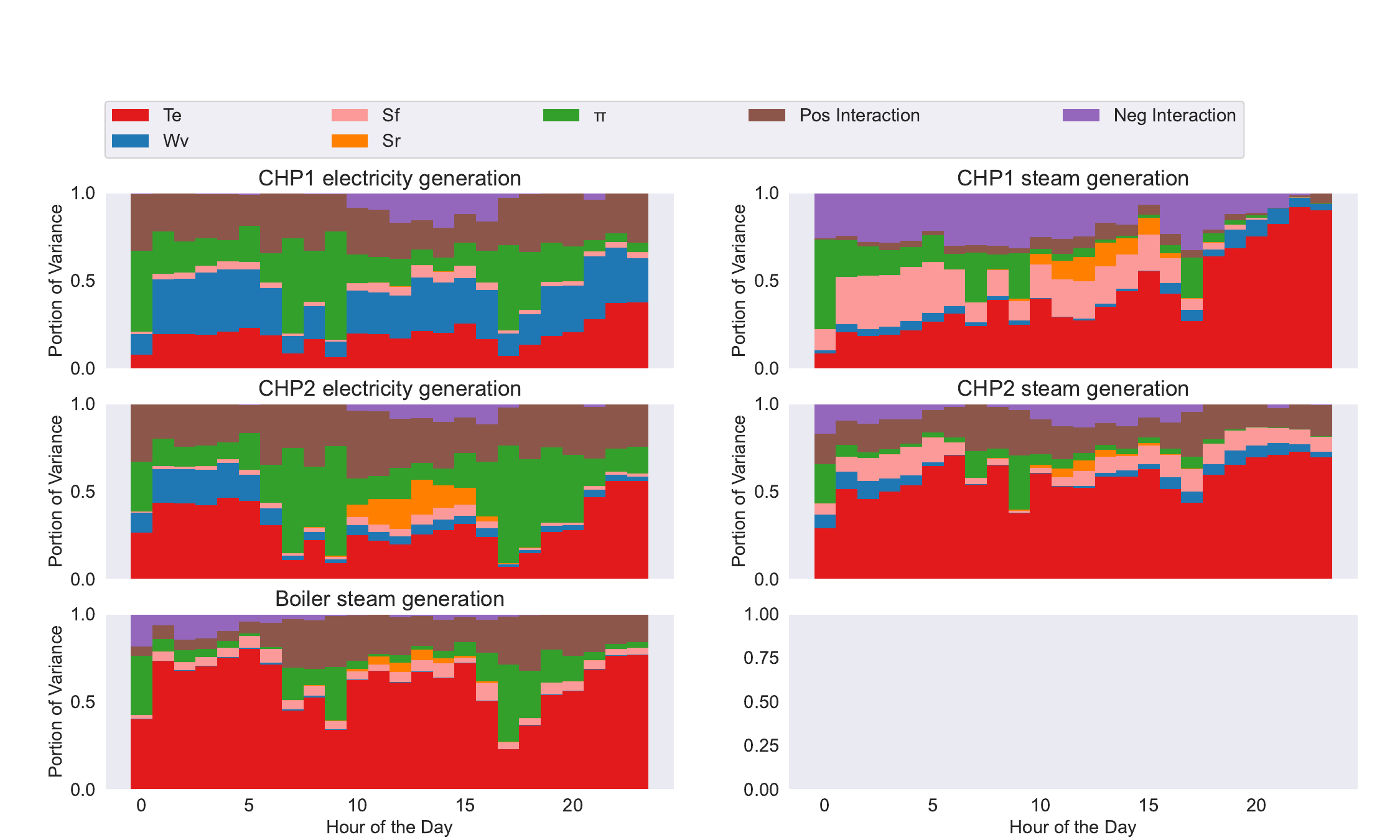}
  \caption{Full first-order TVSA for winter - warm day}\label{fig: fowinterwarm}
  \vspace{-10pt}
  \end{center}
\end{figure}

\begin{figure}[H]
  \begin{center}
  \includegraphics[width=0.9\textwidth]{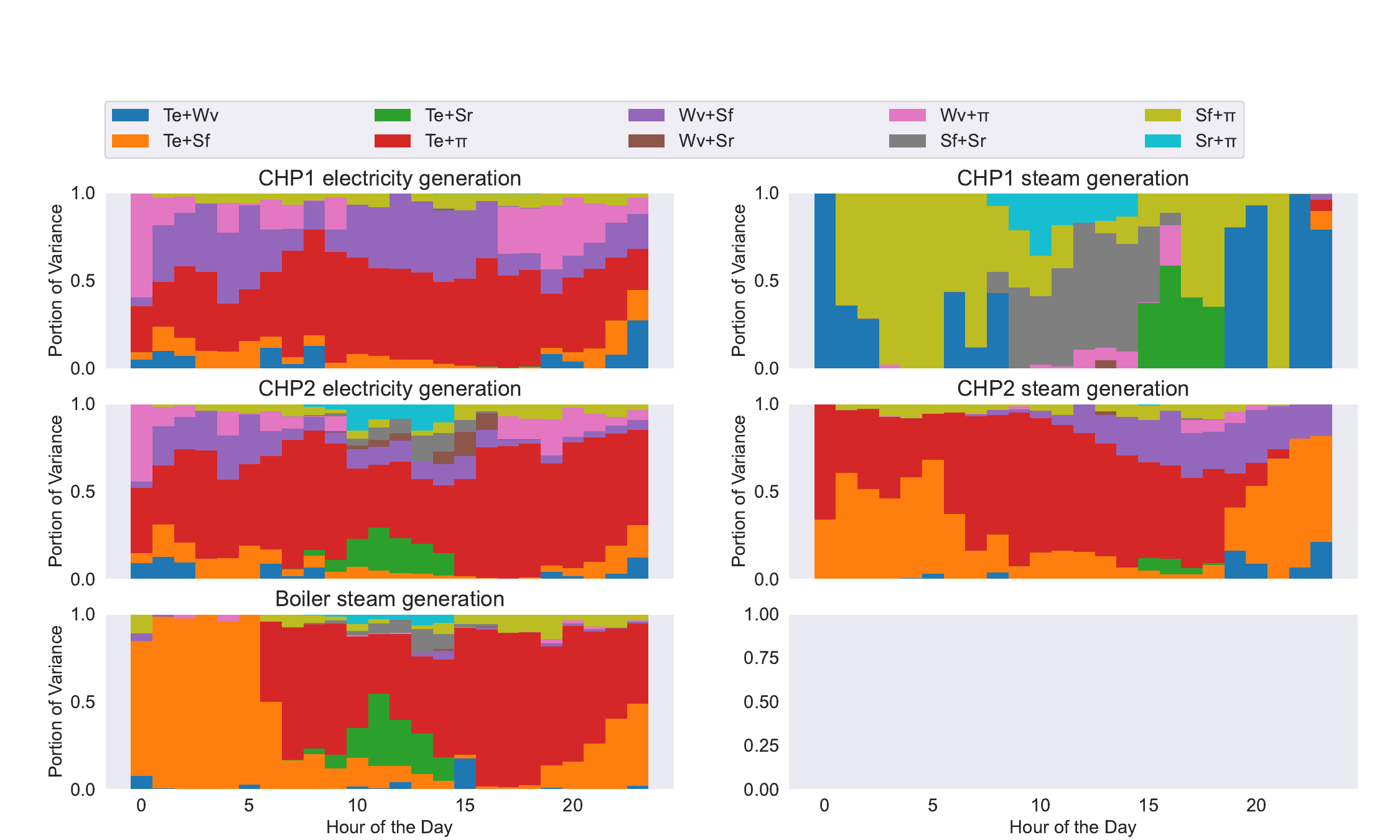}
  \caption{Full second-order positive interaction TVSA for winter - warm day}\label{fig: sowinterwarmpos}
  \vspace{-10pt}
  \end{center}
\end{figure}

\begin{figure}[H]
  \begin{center}
  \includegraphics[width=0.9\textwidth]{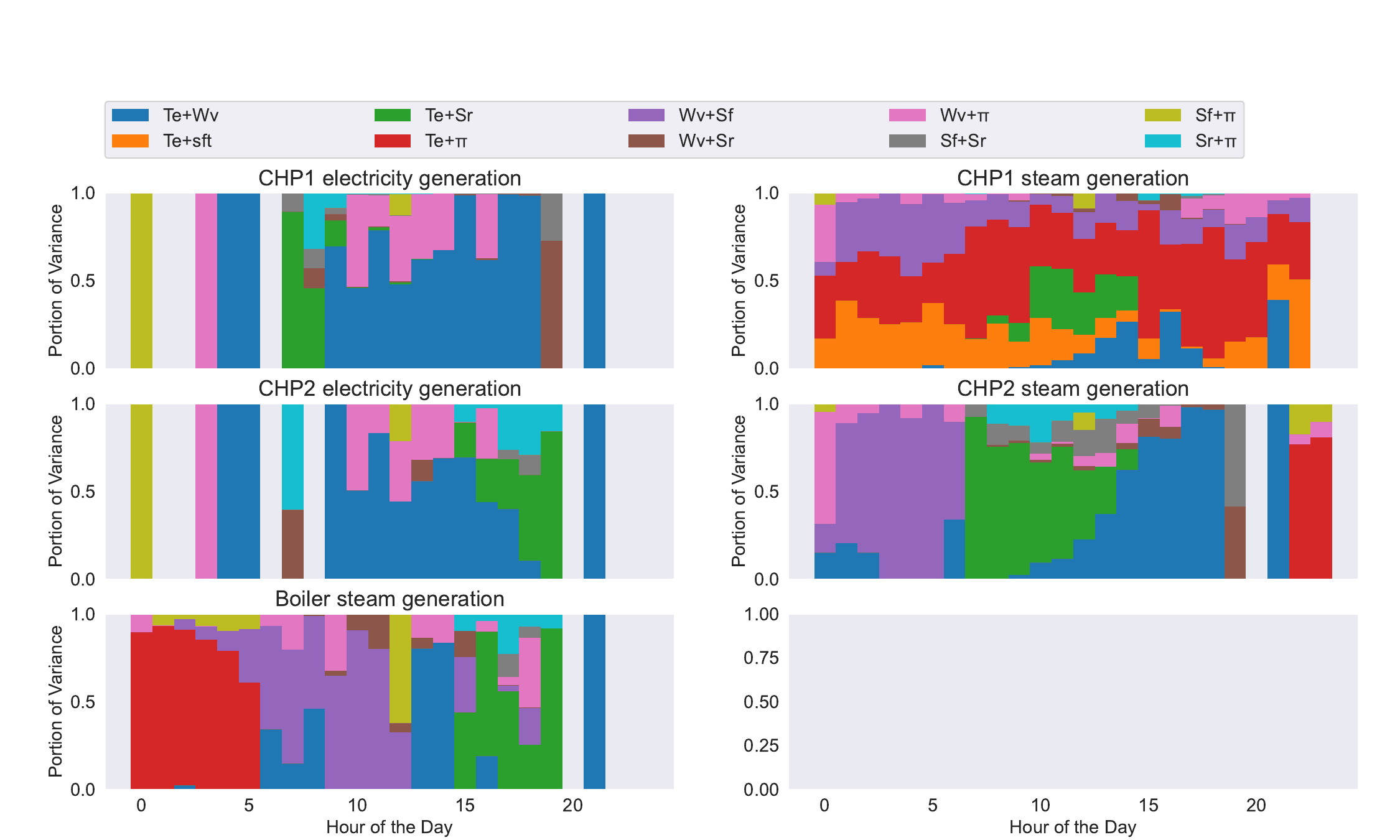}
  \caption{Full second-order negative interaction TVSA for winter - warm day}\label{fig: sowinterwarmneg}
  \vspace{-10pt}
  \end{center}
\end{figure}

\begin{figure}[H]
  \begin{center}
  \includegraphics[width=0.9\textwidth]{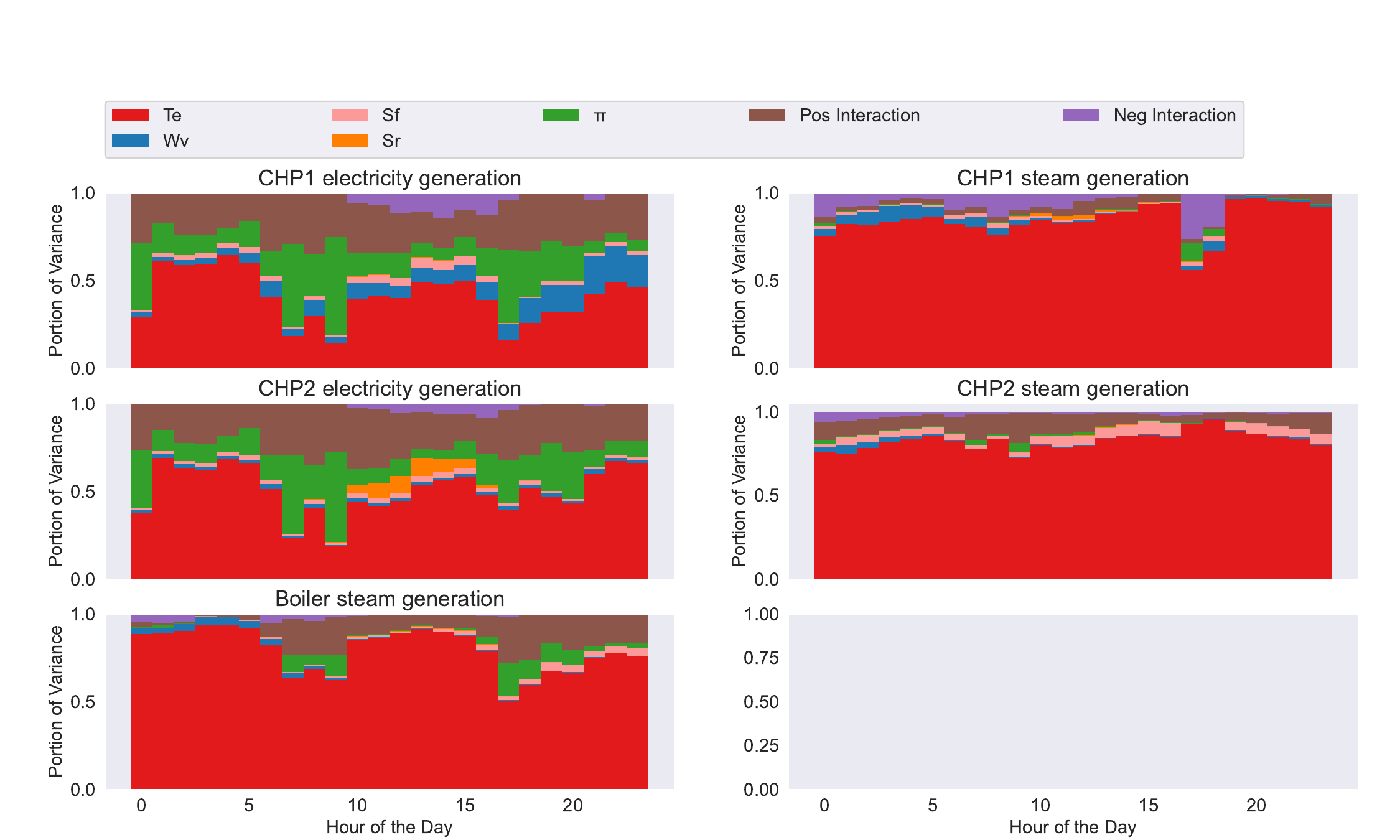}
  \caption{Full first-order TVSA for winter - cold day}\label{fig: fowintercold}
  \vspace{-10pt}
  \end{center}
\end{figure}
\begin{figure}[H]
  \begin{center}
  \includegraphics[width=0.9\textwidth]{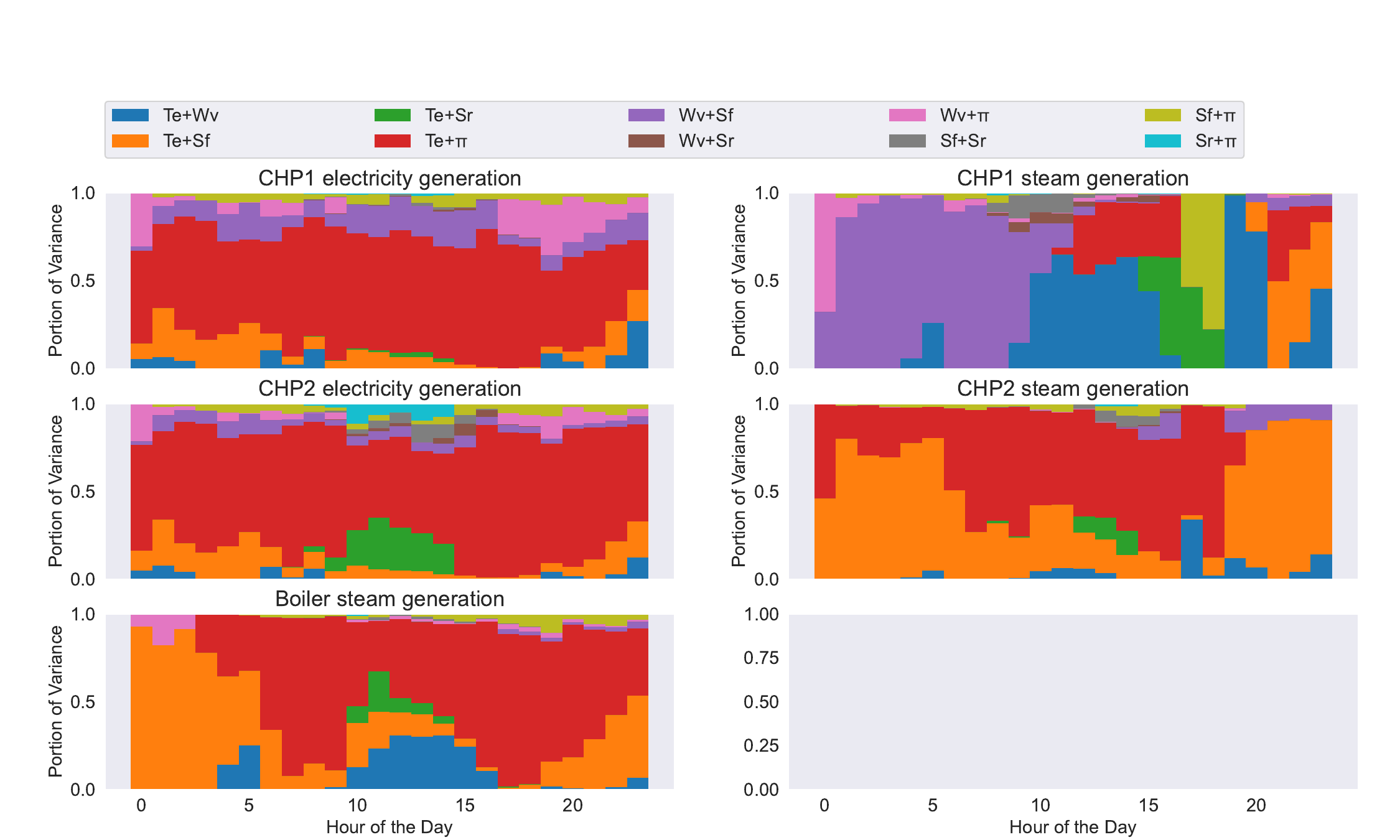}
  \caption{Full second-order positive interaction TVSA for winter - cold day}\label{fig: sowintercoldpos}
  \vspace{-10pt}
  \end{center}
\end{figure}
\begin{figure}[H]
  \begin{center}
  \includegraphics[width=0.9\textwidth]{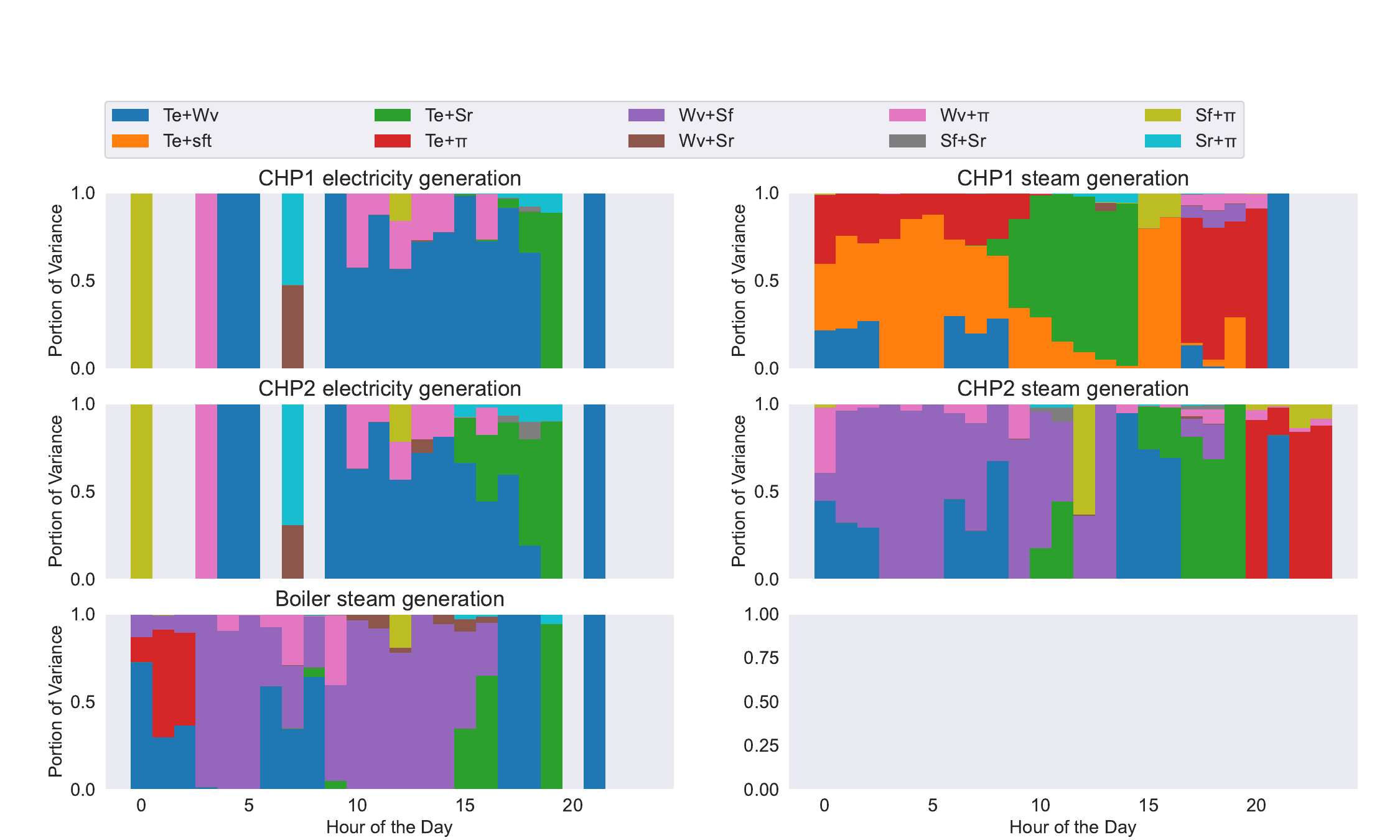}
  \caption{Full second-order negative interaction TVSA for winter - cold day}\label{fig: sowintercoldneg}
  \vspace{-10pt}
  \end{center}
\end{figure}

\begin{figure}[H]
  \begin{center}
  \includegraphics[width=0.9\textwidth]{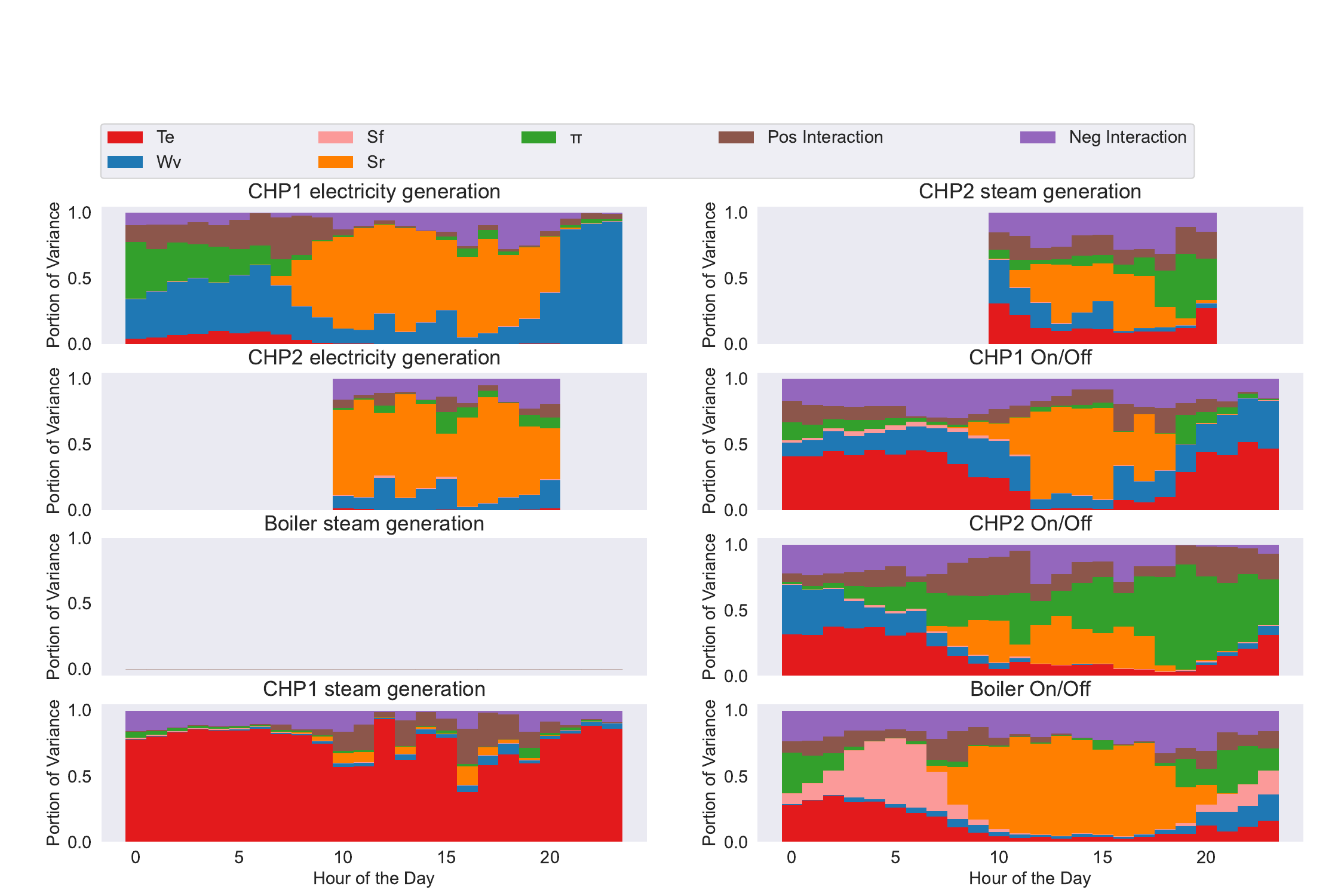}
  \caption{Full first-order TVSA for summer - high price day}\label{fig: fosummerhigh}
  \vspace{-10pt}
  \end{center}
\end{figure}

\begin{figure}[H]
  \begin{center}
  \includegraphics[width=0.9\textwidth]{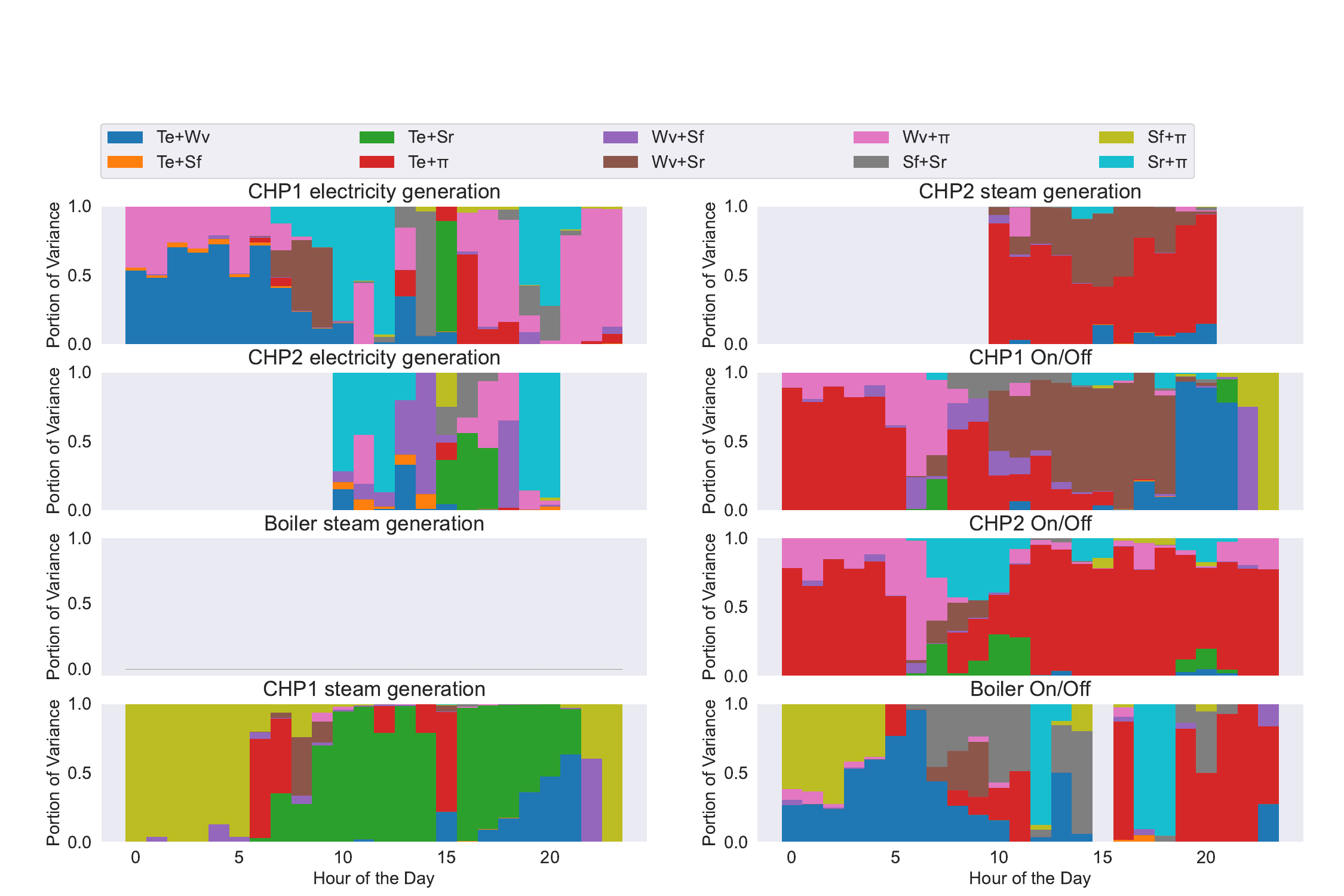}
  \caption{Full second-order positive TVSA for summer - high price day}\label{fig: sosummerhighpos}
  \vspace{-10pt}
  \end{center}
\end{figure}

\begin{figure}[H]
  \begin{center}
  \includegraphics[width=0.9\textwidth]{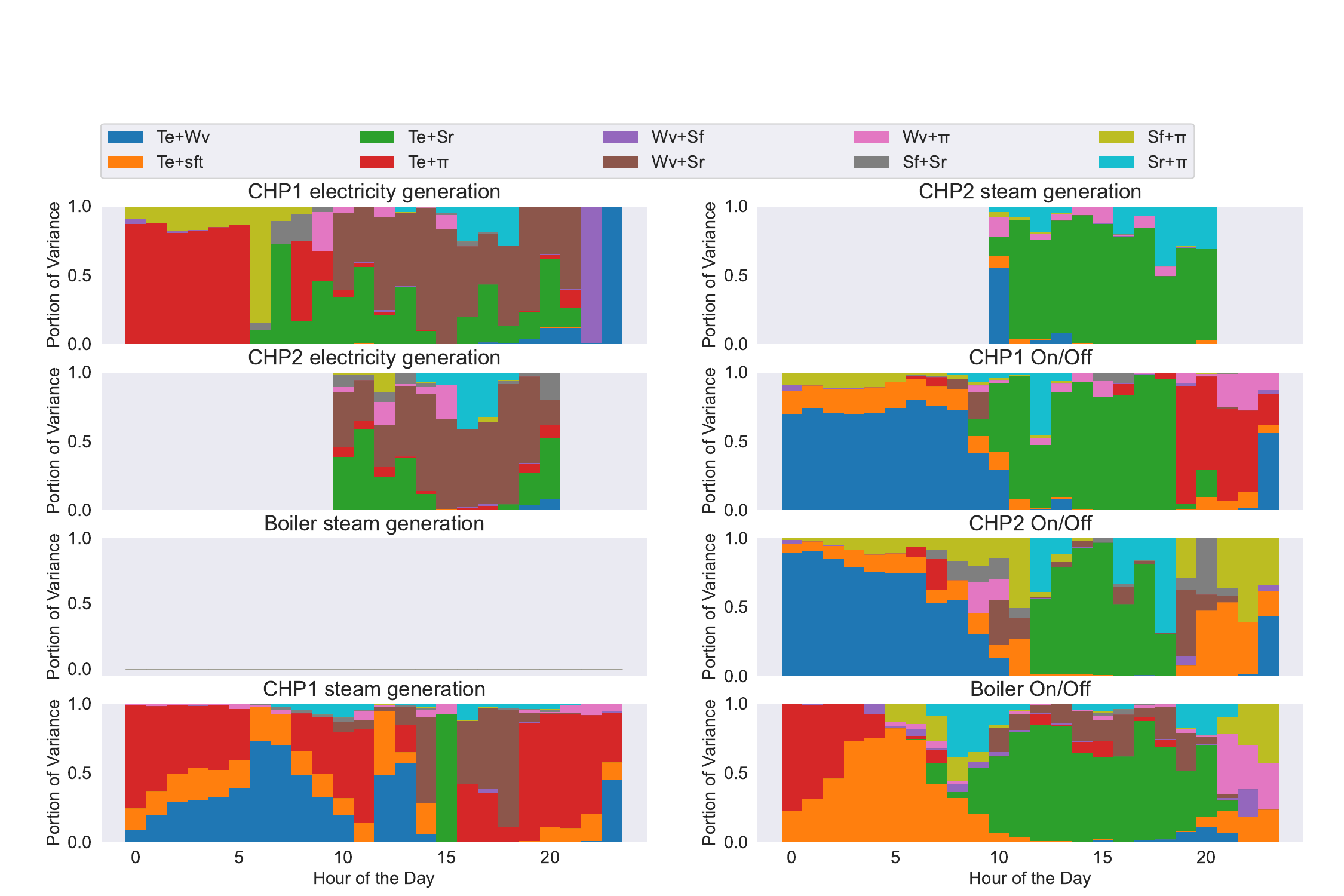}
  \caption{Full second-order negative TVSA for summer - high price day}\label{fig: sosummerhighneg}
  \vspace{-10pt}
  \end{center}
\end{figure}

\begin{figure}[H]
  \begin{center}
  \includegraphics[width=0.9\textwidth]{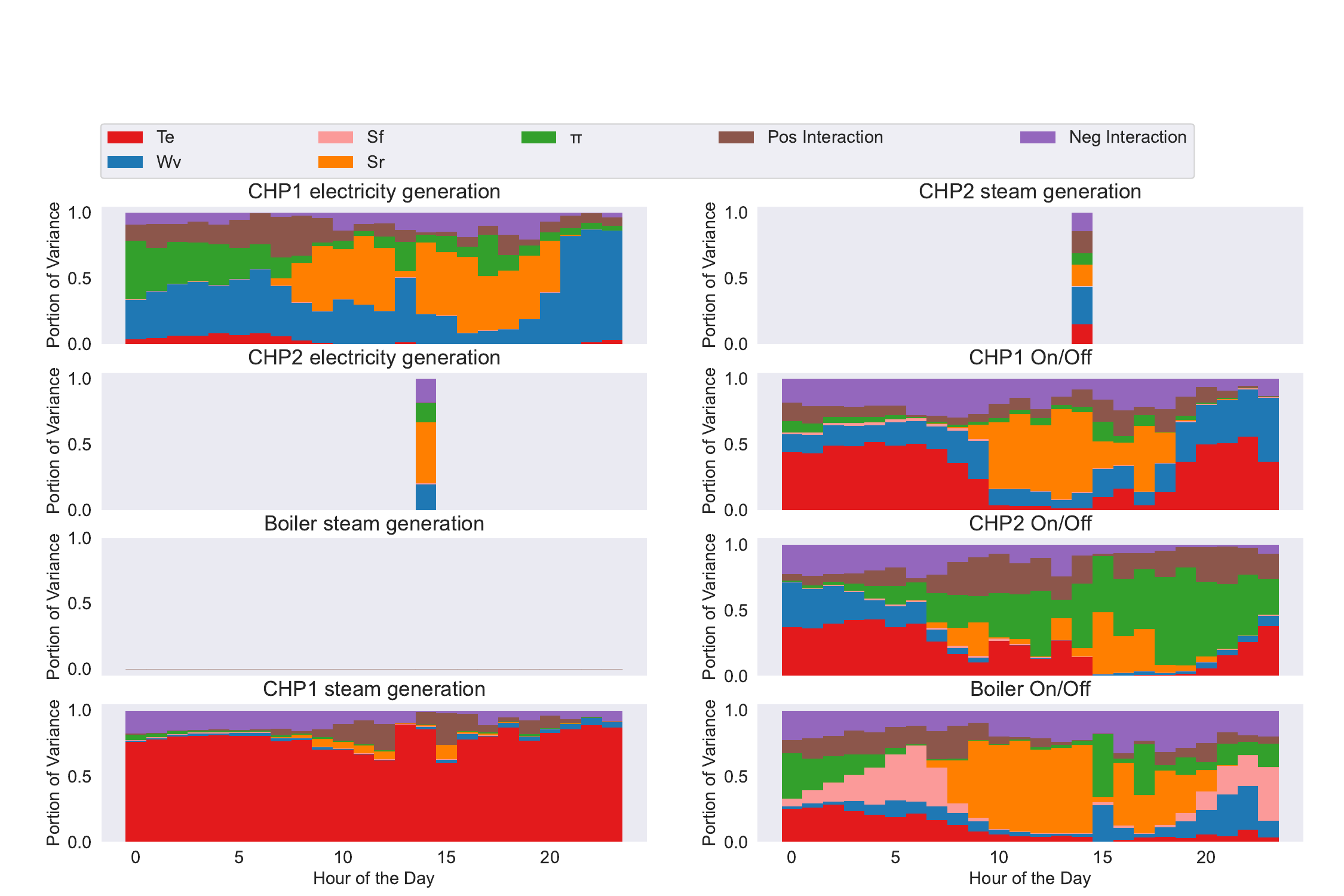}
  \caption{Full first-order TVSA for summer - low price day}\label{fig: fosummerlow}
  \vspace{-10pt}
  \end{center}
\end{figure}

\begin{figure}[H]
  \begin{center}
  \includegraphics[width=0.9\textwidth]{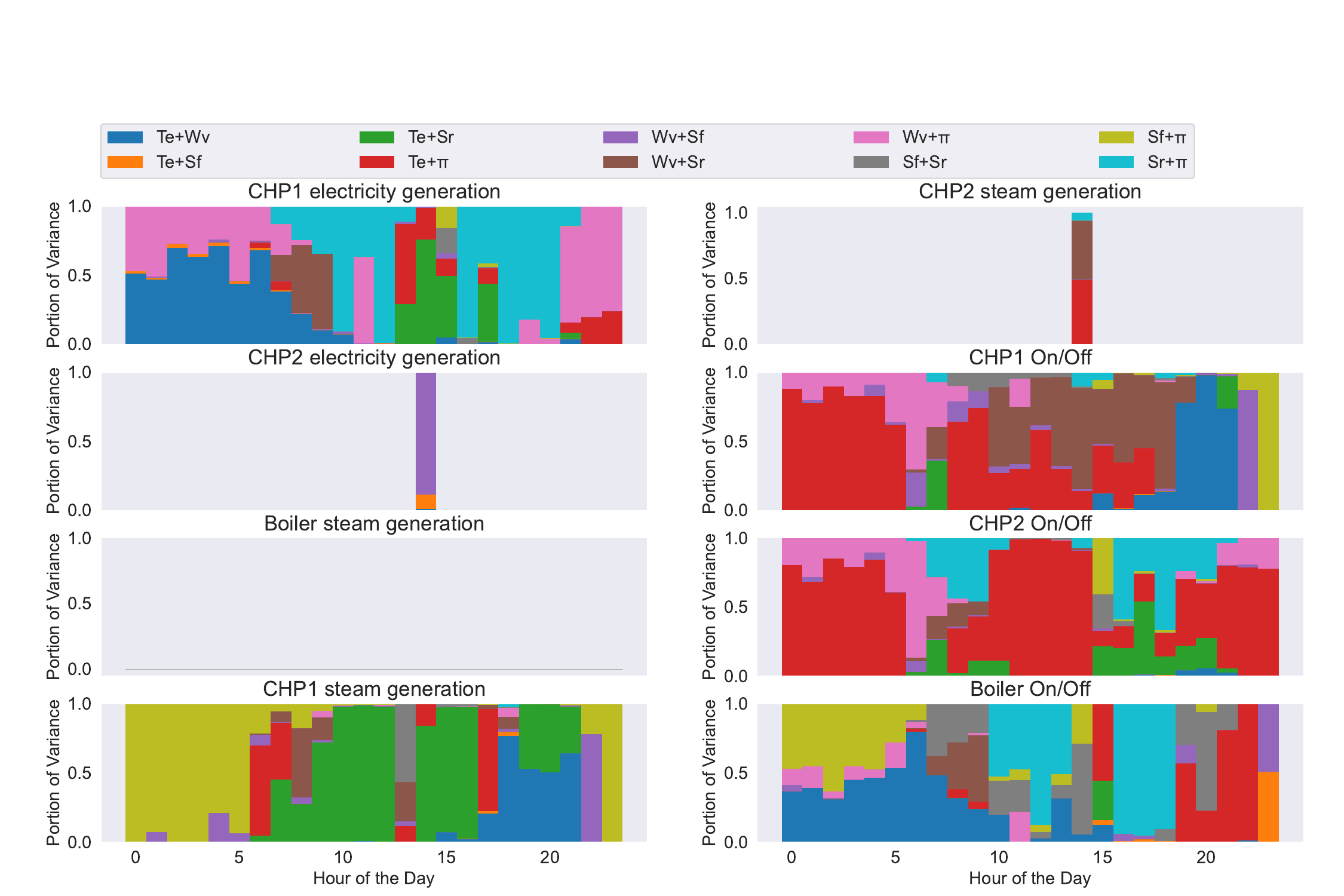}
  \caption{Full second-order positive TVSA for summer - low price day}\label{fig: sosummerlowpos}
  \vspace{-10pt}
  \end{center}
\end{figure}

\begin{figure}[H]
  \begin{center}
  \includegraphics[width=0.9\textwidth]{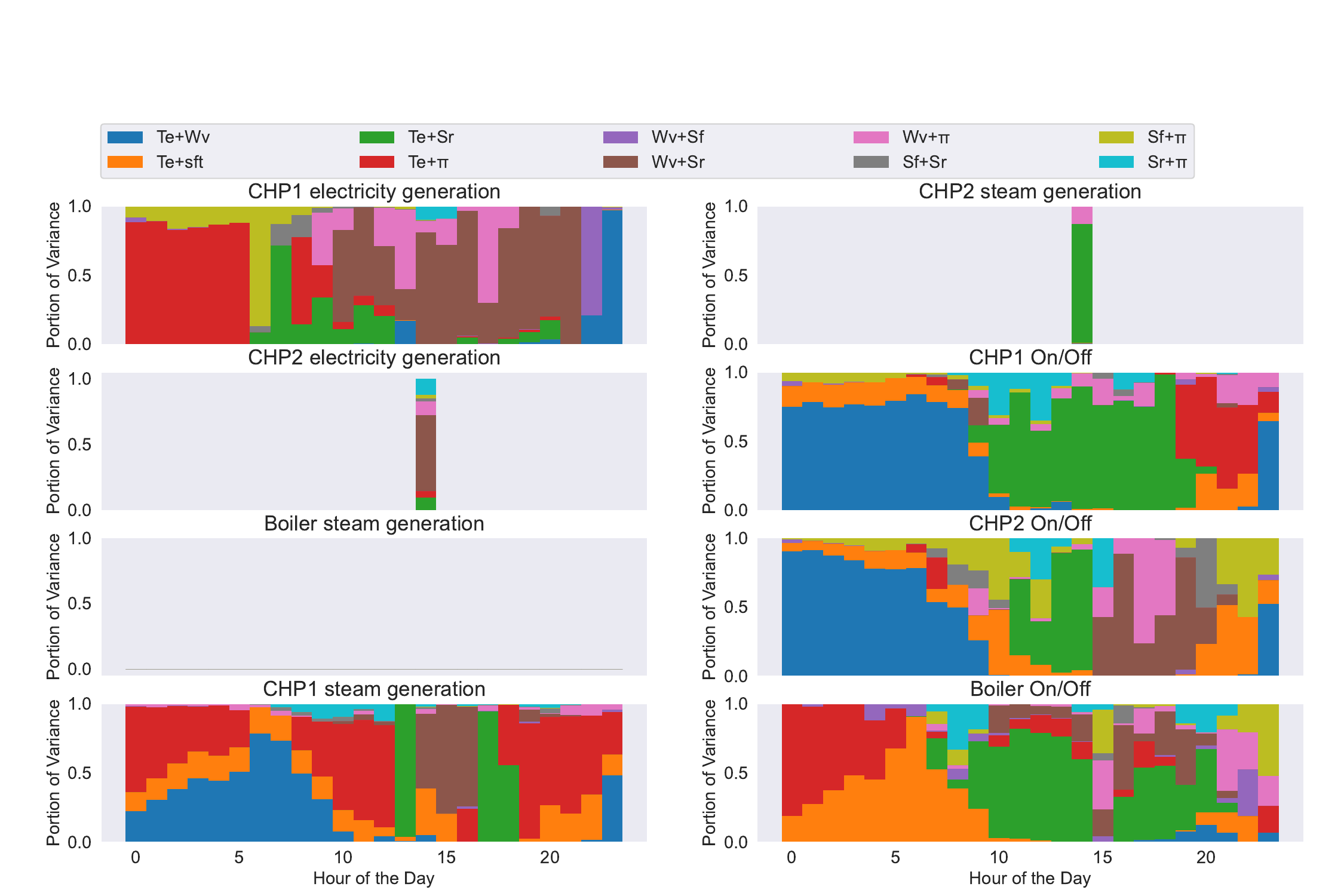}
  \caption{Full second-order negative TVSA for summer - low price day}\label{fig: sosummerlowneg}
  \vspace{-10pt}
  \end{center}
\end{figure}

\bibliography{sample}
